\documentclass[letterpaper,10pt,journal]{IEEEtran}

\usepackage[subtle]{savetrees}
\usepackage{xcolor}
\usepackage{soul}
\usepackage{graphicx}
\usepackage{multirow} 
\usepackage{cite}
\usepackage{relsize}
\usepackage[labelsep=period]{caption}

\usepackage{float}
\floatstyle{plaintop}
\restylefloat{table}

\usepackage{amsmath}
\usepackage{mathtools}
\usepackage{varwidth}
\usepackage{amssymb}

\usepackage{algorithmic}
\usepackage{algorithm}

\usepackage{mathtools}
\usepackage[cmintegrals]{newtxmath}
\usepackage{array}
\usepackage[caption=false,font=normalsize,labelfont=sf,textfont=sf]{subfig}
\usepackage{fixltx2e}
\usepackage{stfloats}
\usepackage{enumitem}
\usepackage{nomencl}
\usepackage[colorlinks=true, allcolors=blue]{hyperref}
\usepackage[numbers, compress]{natbib}
\usepackage{lipsum}

\makenomenclature

\title{\huge{Combating Uncertainties in Wind and Distributed PV Energy Sources Using Integrated Reinforcement Learning and Time-Series Forecasting}}

\begin{document}

\author{\IEEEauthorblockN{Arman Ghasemi*, Amin Shojaeighadikolaei*, Morteza Hashemi}

\IEEEauthorblockA{Department of Electrical Engineering and Computer Science, University of Kansas, Lawrence, KS, USA}
\thanks{*The first two authors contributed equally to this work.}\vspace{-0.7cm}}

\maketitle

\begin{abstract}
% Renewable energy sources are becoming an important part of power system. Wind and solar power generation are among the most popular type of renewable energies. Additionally, development in energy storage system and investigation on integrating energy storage with other types of renewable energies could increase the system performance. However, unpredictable behavior of these sources bring additional complexity and uncertainty to the power system. In this paper, we  propose a framework that combines two machine learning methods in order to run the energy management in the smart grid while addressing the uncertainties because of presence of wind power plant in the generation side as well as active prosumers in the end-user side. In this regard, a deep deterministic gradient method that is a reinforcement learning approach is utilized to apply the energy dispatch in the smart grid. Additionally, long-short term memory as another deep learning method has been used to forecast the wind power output in order to handle the wind uncertainty in the system. Moreover, active prosumers which equipped with solar PV panels and energy storage system  are considered in the residential-user side.
Renewable energy sources, such as wind and solar power, are increasingly being integrated into smart grid systems. However, when compared to traditional energy resources, the unpredictability of renewable energy generation poses significant challenges for both electricity providers and utility companies. Furthermore, the large-scale integration of distributed energy resources (such as PV systems) creates new challenges for energy management in microgrids. To tackle these issues, we propose a novel framework with two objectives: (i) combating uncertainty of renewable energy in smart grid by leveraging time-series forecasting with Long-Short Term Memory (LSTM) solutions, and (ii) establishing distributed and dynamic decision-making framework with multi-agent reinforcement learning  using Deep Deterministic Policy Gradient (DDPG) algorithm. \textit{The proposed framework considers both objectives concurrently to fully integrate them} while considering both wholesale and retail markets, thereby enabling efficient energy management in the presence of uncertain and distributed renewable energy sources. Through extensive numerical simulations, we demonstrate that the proposed solution significantly improves the profit of load serving entities (LSE) by providing a more accurate wind generation forecast. Furthermore, our results demonstrate that households with PV and battery installations can increase their profits by using intelligent battery charge/discharge actions determined by the DDPG agents. 

\vspace{-0.2cm}
\end{abstract}

\section{Introduction}

% Wind power is one of the most important sources of renewable energy, and wind energy is projected to generate 20 percent of U.S. electricity by 2030~\cite{Nrel}.

% \IEEEPARstart{T}{his}
Renewable energy sources (RESs), such as wind and solar, are increasingly being integrated into electric power systems due to their environmental benefits and fuel requirements. According to~\cite{Nrel}, wind power will produce 20 percent of U.S. electricity by 2030. Moreover, the U.S. Energy Information Administration (EIA) reports that solar power installations in the residential sector increased by 34\% from 2.9 GW in 2020 to 3.9 GW in 2021~\cite{EIA}. 
Despite the advantages of wind and solar power for power systems, their increased use poses new challenges to smart grid.

The output of wind power plants (WPPs) is uncertain and highly variable, as it is influenced by atmospheric and climate conditions. As a result of this uncertainty, maintaining a balance between demand and generation can be difficult from the perspective of the \emph{wholesale market}.
Furthermore, the solar photovoltaic (PV) output varies greatly with the weather, ranging from 0\% to 100\%. This high fluctuation makes the system model more complex and uncertain with the increase in residential PV installation. As a result of this additional uncertainty, energy management would be difficult for \emph{retail market} operators.

Independent system operator (ISO) determines wholesale market clearing prices and clearing power quantities in the power system. To hedge against changes in power quantity and price, ISO manages day-ahead (DA) and real-time (RT) markets. The ISO calculates locational marginal prices (LMPs) based on least cost and other system constraints. The uncertainty caused by RESs could lead to a mismatch between supply and demand in the DA and RT markets, which in turn, affects the LMPs. In this situation, the load serving entity (LSE) that is responsible for delivering energy to the end users faces additional price uncertainty other than RESs uncertainty.
Thus, two main challenges need to be overcome from the LSE perspective: (i) managing uncertainty brought on by the presence of RESs, and (ii) managing and distributing energy economically in a setting where there are distributed market participants.

% In recent years, machine learning (ML) and statistical methods have been used extensively to model the uncertainty that results from RESs penetration into the smart grid~\cite{9347810,9145791}. Additionally, to solve the energy management in smart grid, mathematical model-based programming approaches, such as mixed-integer linear programming (MILP), and dynamic and stochastic programming have been widely used~\cite{9223743,rathor2020energy}. Moreover, model-free reinforcement learning (RL) techniques have attracted significant interests in energy management since they do not require an explicit model of the environment~\cite{9398860}. RL and ML algorithms have also been integrated to perform energy management while addressing uncertainty in smart grid applications, such as home energy management~\cite{8681422}.
% However, the problem of integrating RES uncertainty with dynamic energy management by RL, while integrating wholesale and retail market uncertainties that are highlighted in some works~\cite{koltsaklis2020optimization, 6994285,bagheri2022integrating}, remains unsolved.

To address the uncertainty, machine learning and statistical methods have been used extensively~\cite{9347810,9145791}. Additionally, to solve the energy management in smart grid, mathematical model-based programming approaches, such as mixed-integer linear programming (MILP), and dynamic and stochastic programming have been widely used~\cite{9223743,rathor2020energy}. Recently, model-free reinforcement learning (RL) techniques have attracted significant interests in dynamic energy management applications such as home energy management, EV charge controls, and battery optimization since they do not require an explicit model of the environment~\cite{9398860}. 
However, the problem of integrating RES uncertainty with dynamic energy management by RL, while integrating wholesale and retail market uncertainties that are highlighted in some works~\cite{koltsaklis2020optimization,bagheri2022integrating}, is not fully explored yet.

% To incorporate the uncertainty in such dynamic applications, while integrating wholesale and retail market uncertainties that are highlighted in some works~\cite{koltsaklis2020optimization,bagheri2022integrating} have rarely been investigated.

This paper aims to fill this gap by investigating the decision-making problem of the LSE in the presence of uncertainty. In particular, we consider the wholesale and retail markets generation and price uncertainties together with prosumer reactions to the price signal in order to accomplish three objectives: (1) increase LSE profit, (2) reduce prosumers' electricity bills, and (3) decrease peak-to-average ratio of the system.
To achieve these goals, we propose a novel framework that fully integrates Long-Short Term Memory (LSTM) for time-series forecasting and Deep Deterministic Policy
Gradient (DDPG) for taking optimal energy management actions. 
% two machine learning methods, LSTM and DDPG. 
% In this case, LSTM and DDPG operate in harmony to combat uncertainties in wind power generation and prosumers  behavior, respectively. 
LSTM-based time-series forecasting is used to tackle the problem of wind power generation uncertainty. In this case, the DDPG algorithm works in harmony with the LSTM model to establish a distributed decision-making framework that relies on LSTM forecast to optimize the energy management actions. 
% The DDPG algorithm trains agents for the load serving entity (LSE) and active prosumers.
% On one hand, the LSE agent gets the power quantity and marginal prices of the energy producers, as well as available wind energy quantity. Based on the load demand and prosumer willingness to participate in helping the LSE, the LSE agent determines the electricity prices. On the other hand, prosumers' agent tries to take action on their batteries to reduce their electricity bill.
% In order to capture the wind generation uncertainty, another machine learning method using LSTM has been used for the wind power generation forecasting.
Therefore, the main contributions of this paper are as
follows: 
\begin{itemize}
    \item We formulate a two-level optimization problem that considers generation, distribution, and load level simultaneously. The envisioned system model includes wind, solar, and energy storage system as renewable sources to model the generation, consumption and price uncertainty. 
    \item To deal with wind power uncertainty, time-series forecasting is implemented using LSTM module to allow the LSE to increase its profit by dynamically changing electricity prices while taking the LMP uncertainty into account.
    \item Agent-based DDPG reinforcement learning approach is fully integrated with LSTM to run energy management economically through training LSE and prosumers agents.
    \item  We examine the performance of our proposed LSTM-DDPG framework on an IEEE 5-bus system model. Using real wind farm dataset and through extensive numerical results, we demonstrate that the proposed framework effectively enhances the performance of the LSE and prosumers. For instance, with the proposed framework LSE profit increases 86\% compared with having time-of-use pricing scenario.
    % \item The effectiveness of the proposed method is demonstrated using extensive numerical results on IEEE 5 bus system model
\end{itemize}

The rest of this paper is organized as follows. Section~\ref{RelatedWork} presents a summary of related works. The system model and problem formulation are described in Section~\ref{sysModel}.
Section~\ref{DRLframework} introduces the proposed LSTM-DDPG framework, and numerical results are provided in Section \ref{Numerical}. Section \ref{Conclusion} concludes the paper.

\vspace{-0.1in}
\section{Related Work}\label{RelatedWork}
There have been extensive studies on energy management in smart grid applications. Due to the space limitation, we mainly focus on RL-based and deep learning (DL) forecasting works. To be specific, from forecasting perspective, a growing body of literature have examined the forecasting using deep learning  to address uncertainty in smart grid~\cite{9347810,9145791,9395437,7885096,8039509}, and from decision-making perspective, RL has been extensively utilized in smart grid energy management applications in the presence of uncertainties~\cite{8742669,li2020real,samadi2020decentralized,guo2022real,9106862,8331897,dolatabadi2022novel,lei2020dynamic}. However, limited works have addressed the energy management by leveraging joint RL-based decision-making and DL-based forecasting framework.
% In this section, we present an overview of the related works, and highlight the main contributions of our work compared with the prior studies. 
% To this end, we consider related works that examine deep learning methods to address the uncertainty. In addition, the works that utilized RL methods to solve the energy management in smart grid and combination of DL and RL are explored.

\textbf{(1) Learning-based forecasting.} Among deep neural networks, different variants of recurrent neural networks (RNNs) such as LSTM have attracted much interest from the research community because of their internal memory features~\cite{9347810,8039509,9145791}. For instance, \cite{9347810} proposes a stacked DL model based on RNN variants for both renewable energy and electricity load prediction that requires fewer parameters to train.
The authors in \cite{9145791} discuss uncertainty modeling problems in smart grid, as well as proposing a method of clustering combined with an LSTM model to make electricity load and price predictions more accurate. 
In~\cite{8039509}, an LSTM-based framework is proposed for forecasting individual residential users' electric load, which is compared with system aggregated load forecasting.
These works are mostly focused on handling uncertainties using forecasting techniques, while not considering energy management and dynamic decision-making under such uncertain conditions. 

\textbf{(2) Reinforcement learning for energy management.} Reinforcement learning has recently become popular in handling energy management problems. For instance, in~\cite{8742669}, double deep Q-learning Network (DDQN) method was used to manage a community battery energy storage system (BESS). To combat the price signal uncertainty, they considered $\pm 5\%$ uncertainty interval. Likewise in~\cite{li2020real}, the authors presented a demand response framework of scheduling home appliances using deep RL method. They utilized model predictive control (MPC) to forecast the future electricity price and outdoor temperature. In another work~\cite{samadi2020decentralized}, the authors utilized Rayleigh and Beta probability distributions for modeling renewable uncertainty in multi-agent Q-learning framework for micro grid (MG) energy management.

Unlike DQN, which has discrete action space, policy-gradient methods such as proximal policy optimization (PPO)~\cite{guo2022real} and DDPG~\cite{9106862,8331897,dolatabadi2022novel,lei2020dynamic} work in continuous action spaces and provide a more realistic control scheme. For instance, the work in~\cite{guo2022real} and \cite{dolatabadi2022novel} utilized common statistical Weibull and Beta distributions for wind speed and solar irradiance prediction in PPO and DDPG framework, respectively. In~\cite{lei2020dynamic}, the authors utilized DDPG with finite-horizon partially observable Markov decision process (POMDP) model to capture the future electricity consumption and PV generation in MGs energy management. It is worth noting that none of these works~\cite{8742669,li2020real,samadi2020decentralized,guo2022real,9106862,8331897,dolatabadi2022novel,lei2020dynamic} considered DL forecasting frameworks.

% Energy management is regarded as a core part of smart grid to effectively coordinate energy sharing among energy sources and loads. RL methods have recently become popular frameworks to handle energy management with uncertainty due to the increasing use of renewable energy sources~\cite{8742669,guo2022real,9106862,8331897,9016168}. For instance, the authors in~\cite{8742669} proposed a double deep Q-learning Network (DDQN) method to manage a community battery energy storage system (BESS).

%  Furthermore, DQN has a discrete action space, unlike policy-gradient methods such as PPO~\cite{guo2022real} and DDPG~\cite{9106862,8331897} that has a continuous action space that is more realistic model to capture various quantities in smart grid~\cite{9106862,8331897}.
% In \cite{9106862}, the electricity market was modeled and a bidding problem between generation companies was solved using DDPG. In \cite{8331897}, a multi-agent distributed energy management system using Q-learning framework with time-of-use pricing was proposed by the authors for a microgrid energy market with a centralized battery pack and renewable energy sources. There was no consideration of uncertainty in either of these two works.

\textbf{(3) Combined forecasting and energy management.} Combined forecasting and RL framework has been used in several applications such as route trajectories and autonomous vehicle controls~\cite{rasheed2020deep}. This technique is proven to be efficient in smart grid energy management applications~\cite{8681422,9684239,xu2020multi,9061038,8521585,9163332}. The authors in~\cite{8681422} utilized a forecasting model based on artificial neural network (ANN) to predict future price for home energy management. In~\cite{9684239}, the authors developed  a two-level RL framework to deal with the optimal pricing of multiple MGs. To combat the uncertainties, they utilized prediction interval using neural network with bootstrap. Robustness is their main focus that makes their solution applicable for worst-case scenarios. In the other work~\cite{xu2020multi}, an Extreme Learning Machine (ELM) based feedforward NN was used for predicting the future trend of electricity price and PV generation. This framework was integrated with a DQN framework for optimizing a home energy management problem.
Our approach differs from these works~\cite{8681422,9684239,xu2020multi} in that we combine LSTM engine, which works well with processing time-series forecasting data and DDPG, which is applicable and more realistic for continuous spaces.
% It is also worth noting that these works~\cite{8681422,9684239,xu2020multi} are different from ours, while we combined LSTM  engine, which is well suited for processing continuous time-series data generated by wind generation, and DDPG, which is applicable and more realistic for continuous spaces.
In addition, our work is an economic-oriented framework applicable for all scenarios. 

Recently, LSTM has been integrated to RL frameworks in several researches~\cite{9061038,8521585,9163332}. The authors in~\cite{9061038,8521585} combined LSTM with single DQN agent for optimizing battery energy arbitrage and EV charge/discharge scheduling, respectively. Unlike the work~\cite{8521585}, the authors, in~\cite{9163332}, combined a single DDPG agent with LSTM engine to continuously control the EV charge/discharge trend. In all of these three works, LSTM is utilized to predict the future electricity price to combat price uncertainty. 
% Although we borrow the idea of combining LSTM and DDPG in~\cite{9163332}, we modify it for MARL setting and leverage it across electricity market.

It can be seen from the aforementioned literature that the existing research has made an in-depth discussion on home energy management, battery optimization, multi MGs, and EV charge/discharge control in the presence of uncertainties. However, to the best of our knowledge, integrating wind uncertainties with dynamic energy management using RL methods in a unified framework has not been investigated before.
Therefore, in this paper, we develop a framework based on DDPG and LSTM to model the wholesale and retail markets simultaneously and consider the uncertainty of RES and price for (1) wind power generation, (2) LMP uncertainty, (3) retail price, and (4) demand side uncertainty in the presence of prosumers.

\section{System Model and Problem Formulation}\label{sysModel}
In this section, we present the system model, followed by the problem formulations for all market players (i.e., the LSE and distributed prosumers).
\vspace{-0.1in}
\subsection{Power Market Model}
\begin{figure}[!t]
\centering
    \includegraphics[scale=0.20, trim =0.8cm 0.5cm 0.8cm 0.9cm, clip]{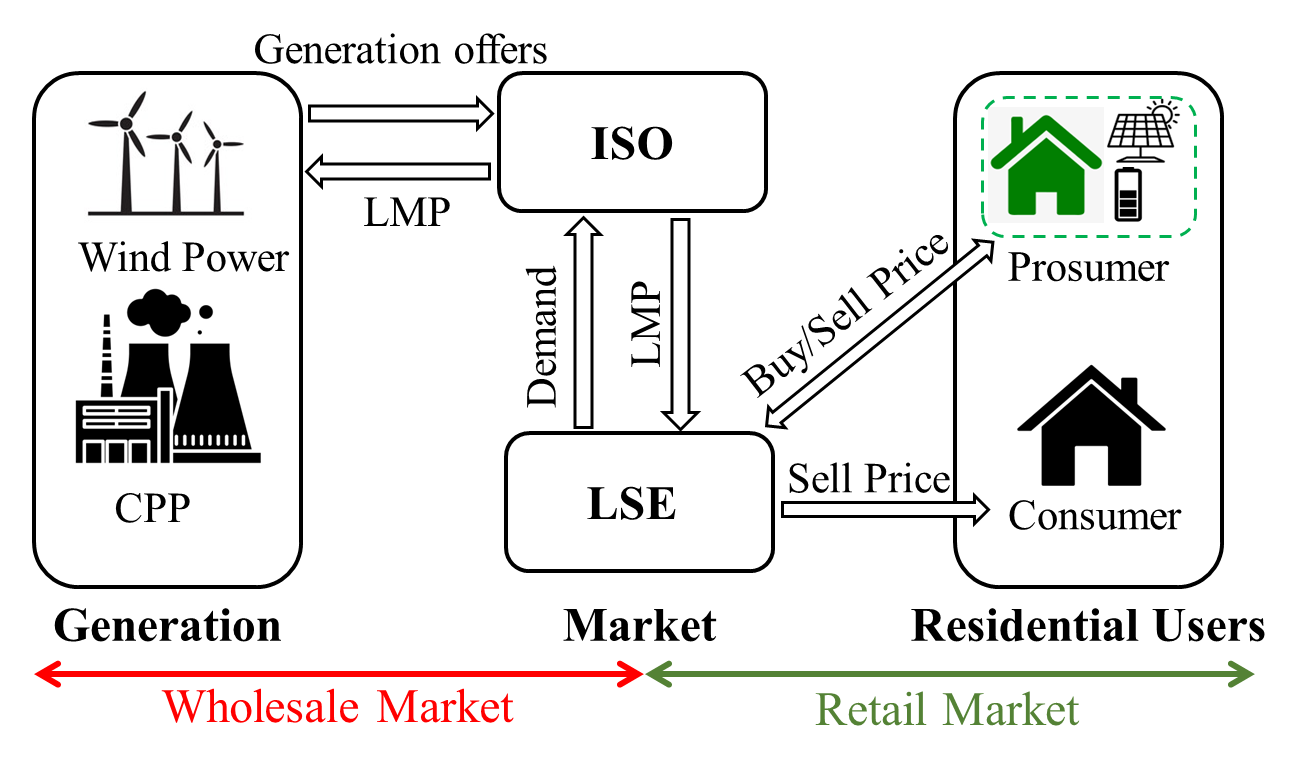}
    \caption{Electricity market structure with  prosumers and wind power plant. 
  }
    \label{Model}
    \vspace{-.15in}
\end{figure}

% As a result of integrating renewable generation into the distribution-level energy market,
This study develops a decision-making framework that combats uncertainties associated with renewable generation penetration, while optimizing the economic benefits for electricity providers and residential users. To this end, the envisioned system model is illustrated in Fig.~\ref{Model}. The model includes active prosumers as distributed market players, equipped with PV rooftop panels and energy storage systems. LSE represents a distribution utility player, which is responsible for aggregating load on behalf of residential users and making appropriate arrangements in wholesale markets to meet the total load. An ISO is responsible for ensuring reliability and adequacy of the power system. Wind power plant (WPP) and conventional power plant (CPP) are the two types of large-scale generation units.

% Our envisioned energy market consists of day-ahead (DA) and real-time (RT) markets, where ISO determines the clearing price and clearing power quantities for both markets. In this paper, we leverage the LMPs calculation framework provided in~\cite{fang2016strategic}. The DA and RT LMPs calculations are following the same rules.
% Uncertainty in wind generation affects LMP directly and causes uncertainty for determining LMP. 
% For the LSE, this means a shift of the uncertainty from wind generation to LMPs. From retail market perspective, this uncertainty has a direct impact on electricity prices determined by LSE. Consequently, retail market players, such as prosumers, will encounter an increase in their electricity bills.

% From the wholesale market perspective, wind power generation uncertainty causes a mismatch between DA predictions and RT available power, as shown in Fig.~\ref{supply}. This figure illustrates the sorted incremental cost curve of a piecewise linear approximation when various power producers are present.
% According to the figure,
% wind power uncertainty affects the LMP that increases the risk of monetary loss for all the participants in both wholesale and retail markets. 

\begin{figure}[!t]
  \centering
    \includegraphics[scale=0.32, trim = 0.3cm 0.2cm 0.5cm 0.9cm, clip]{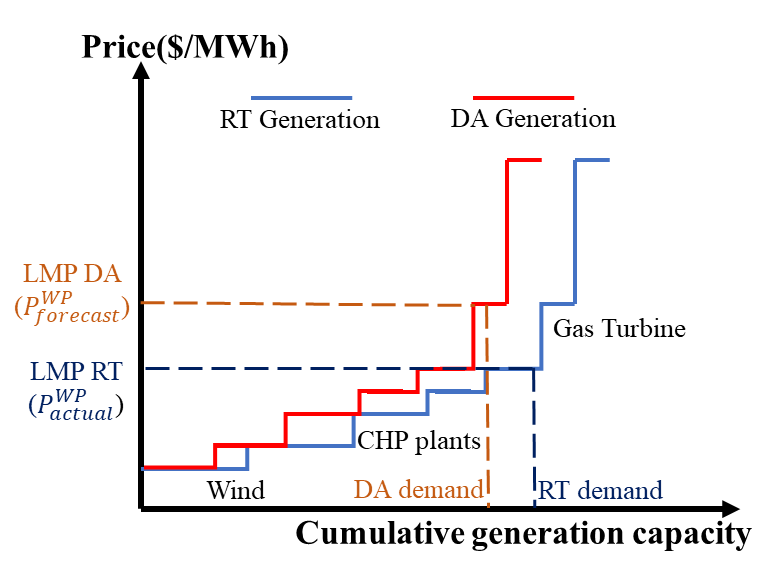}
    \caption{Marginal electricity producer cost curve.}
    \label{supply}
    \vspace{-.18in}
\end{figure}

Our envisioned energy market consists of day-ahead (DA) and real-time (RT) markets, where ISO determines the clearing price and clearing power quantities for both markets. In this paper, we leverage the LMPs calculations framework provided in~\cite{fang2016strategic}. The DA and RT LMPs calculations are following the same rules. From the wholesale market perspective, uncertainty in wind generation affects LMPs in both DA and RT markets directly. Consequently, this uncertainty causes a mismatch between DA predictions and RT available power, as shown in Fig.~\ref{supply}. This figure illustrates a sorted incremental cost curve of a piecewise linear approximation when various power producers are present. As illustrated, this mismatch increases the risk of monetary loss for wholesale participants, e.g. LSE. From the retail market perspective, LSE dynamically determines the retail electricity market to overcome its monetary loss. Consequently, it will have an adverse effect on retail market players' electricity bill, such as prosumers. As a result, we have a two-level optimization problem. On one hand, the LSE participates in wholesale market to procure the energy needs for residential users, while trying to maximize its local profit. On the other hand, in the residential users level, active prosumers aim to reduce their electricity bills by participating in the retail market. Next, we define this two-level optimization problem.

\vspace{-0.1in}
\subsection{Distributed market players}
In the residential-users network,  prosumers participate in the retail market as distributed players.
% Solar PV panels and energy storage system (ESS) are considered as a distributed energy resource for active prosumers. 
The prosumers will produce energy to meet their demand, charge their battery, or sell the energy back to the grid during the peak PV generation times or high price periods. Each user $i$ belongs to the set of active prosumer (denoted by $\mathcal{N^P}$) or the set of passive consumers (denoted by $\mathcal{N^C}$). Thus, $ i \in \mathcal{D} = \left\{ \mathcal{N^P} \cup \mathcal{N^C} \right \}$. The active prosumers' energy demand at the time slot $t$ is defined as follows:
\vspace{-0.05in}
\begin{equation}\label{eq:pro}
e_{i,t} = d_{i,t}-g_{i,t}-b_{i,t},\:\: \forall i \in \mathcal{N^P},
\end{equation}
where $d_{i,t}$, $g_{i,t}$, and $b_{i,t}$ denote the energy consumption, PV generation, and energy charge/discharge from the battery, respectively. 
PV generations must be limited as follows:
% by their maximum generation capacity, i.e.:
\vspace{-0.05in}
\begin{equation}\label{eq:PV constraint}
  0\le\ g_{i,t} \le g^{\text{max}}_{i},\: \forall i \in \mathcal{N^P}.
\end{equation}

% \textbf{Energy Storage System (ESS):} 
The prosumer's battery is modeled by the general ESS characterization such as, charging and discharging rate, maximum capacity, and state of charge of the battery, which are described as follows:
\vspace{-0.05in}
\begin{equation}\label{eq:batttery_limit}
 SoC^{{\min}}_{i} \le \frac{1}{{{Q_i}}}\sum_{t\in T}{b_{i,t}} + So{C_{{i}}}(0) \le SoC^{{\max}}_{i},\: \forall i \in \mathcal{N^P,} 
\end{equation}
\begin{equation}\label{eq:battery charge}
  b^{dis,min}_{i} \le b_{i,t}\le b^{charge,max}_{i},\: \forall i \in \mathcal{N^P}.
\end{equation}

In addition, each end user's total load $L_{i,t}$ for consumer and prosumer is defined in Eq.~\eqref{eq:demand}, i.e.,
\vspace{-0.05in}
\begin{equation}\label{eq:demand}
    {L_{i,t}}=\begin{cases}
        d_{i,t} & \text{if} \; i \in \mathcal{N^C}\\
    e_{i,t} & \text{if} \; i \in \mathcal{N^P}. 
            \end{cases}
\end{equation}

Thus, the aggregated load that represents the energy transfer between the LSE and residential users is obtained as $L^D_t =  \sum_{i=1}^{{|\mathcal{D}}|}{L_{i,t}}$.

% ${E_{t}}=\sum_{i = 1}^{|\mathcal{N^P}|} (1-\Tilde{\lambda}_i)|e_{i,t}|$.

The goal of each prosumer is to maximize its local revenue, or conversely, minimize its total electricity bill calculated as follows:
\begin{equation}\label{eq:prosumerobj}
\mathcal{E}_i^{pro} = \sum_{t \in T} \left \{ (1-\Tilde{\lambda}_i) e_{i,t}C^s_t + \Tilde{\lambda}_i e_{i,t}C^b_t \right \} ,\:\: \forall i \in \mathcal{N^P}, 
\end{equation}
in which $\Tilde{\lambda_i} \in \{0,1\}$, such that if $e_{i,t}$ is non-negative, then $\Tilde{\lambda}_i = 1$. This condition implies that the prosumer $i$ needs to buy the energy from the LSE at the price of $C^b_t$. On the other hand, $\Tilde{\lambda}_i = 0$ means that the prosumer sells its excess energy back to the LSE at the price of $C^s_t$. 

Given the formulated energy exchange among distributed users and LSE, the objective of  users is to reduce their electricity bills. To this end, prosumers decide to take an action on their batteries after receiving the electricity price from the LSE. In this regard, the optimization problem in the residential level is defined as follows:
\vspace{-0.05in}
\begin{equation}
 \textbf{User level:} 
 \begin{cases}
 \mathop{\mathrm{minimize}}\limits_{b_{i,t}} & \sum\limits_{i\in\mathcal{N^P}}\mathcal{E}_i^{pro},   \\ 
    \text{subject to:} & \eqref{eq:pro}  \textbf{\&} \eqref{eq:PV constraint} \textbf{\&} \eqref{eq:batttery_limit} \textbf{\&} \eqref{eq:battery charge}
    \textbf{\&} \eqref{eq:demand}.  
\label{eq:UserLevel}
\end{cases}
\end{equation}

It is also noteworthy that the aforementioned constraints determine prosumers' power balance equation, PV generation limit, as well as battery charging and discharging limits. The control variable in this optimization problem, denoted by $b_{i,t}$, is the amount of energy that is charging or discharging from the prosumers' battery.

\subsection{Load Serving Entity}
The LSE seeks to minimize its own total profit, while incorporating different uncertainty sources, ranging from wind power generation, PV rooftop panels, and net demand. The total cost (TC) of the LSE is defined as follows:
\begin{align}
    \begin{aligned}
        TC = \sum_{t\in T} \left\{ {L^{DA}_t {\rho}^{DA}_t} +\sum\limits_{i = 1}^{|\mathcal{N^P}|} e_{i,t}C^b_t \Tilde{\lambda_i} + \Delta L_t {\rho}^{RT}_t \right\}.\label{eq:CostLSE}
    \end{aligned}
\end{align}

The TC incorporates the total cost of the LSE to procure energy from the wholesale market as well as from the distributed market players, i.e., active prosumers. $L^{DA}$ and $\rho^{DA}$ indicate the forecasted load and LMP in DA market. $e_{i,t}$ indicates the demand from the $i^{th}$ prosumer.
% and $C_t^b$ is the electricity buy price determined by the LSE.
$\Delta L_t =  L^{RT}_t - L^{DA}_t$ represents the RT and DA demand deficiency, which is used to calculate the LSE cost for the \emph{load uncertainty} in the RT market. Thus, the LSE net profit can be calculated by subtracting the LSE costs from the aggregated residential load demand, i.e.: 
\begin{align}
    \begin{aligned}
        \mathcal{R}^{LSE} = \sum_{t\in T} \left\{ \sum\limits_{j=1}^{|\mathcal{N^C}|}{L_{j,t}^{RT}C_t^s} + \sum\limits_{k=1}^{|\mathcal{N^P}|}{L_{k,t}^{RT}C^{s}_t} \right\} - TC. \label{eq:RevLSE}
    \end{aligned}
\end{align}

% \subsection{Power Balance and Energy Costs}
% The ISO/RTO is responsible for maintaining the balance between demand and generation, as shown in Figure. \ref{Model}. The ISO collects the demand and generations from the LSE and power producers, respectively. Accordingly, the system marginal price and the generators' power commitment will be determined for the whole system.
% For both DA and RT markets, LMPs are calculated. 
% Power balance is a major factor in determining LMP, and LMP is determined based on the cost of a generator that could clear the market.
% The impact of wind power uncertainty on the LMP price is also taken into account.
\textbf{Power Balance:} LSE operations should maintain balance between consumption and generation at each time instants. 
Therefore, power balance equation is defined as follows:
\begin{equation}\label{eq:Powerbalance}
  \sum\limits_{i = 1}^{|\mathcal{D}|} L_{i,t}   = \sum\limits_{i=1}^{|\mathcal{N^G}|} P_{i,t}^{G} + P^{wp}_t,
\end{equation}
where the aggregated demand of the residential users in the left-hand side should be met by the power producers $ P_{i,t}^{G}$, as well as wind power generator that is indicated as $P^{wp}_t$.

\textbf{Generation Constraints:} According to Fig.~\ref{Model}, conventional power plants are considered as one type of generator, along with wind power plant as the renewable energy source in generation side. 
% A wind power plant is considered for assessing the uncertainty caused by intermittent wind generation.
Each generation facility $i \in \mathcal{N^G}$ has a minimum and maximum generation limit, as described in Eq.~\eqref{eq:GenLimits} and Eq.~\eqref{eq:Wind constraint}. The generation ramp rate is defined in Eq.~\eqref{eq:GenRamp} to demonstrate how quickly a plant can change its output. Therefore, we have: 
\begin{equation}\label{eq:GenLimits}
  P_i^{G,\min} \le P_{i,t}^G \le P_i^{G,\max}, \forall i \in \mathcal{N^G},
\end{equation}
\begin{equation}\label{eq:Wind constraint}
  0\le\ P_t^{wp} \le P^{wp,\text{max}},
\end{equation}
\begin{equation}\label{eq:GenRamp}
  RR_i^{\min} \le P_{i,t+1}^G - P_{i,t}^G \le RR_i^{\max}, \forall i \in \mathcal{N^G}.
\end{equation}

% For LSE level, a maximization problem is defined based on the power balance and generation constraints.

\textbf{LSE Optimization:} In the upper-level market, LSE aims to maximize its own profit. LSE is trying to achieve this objective through changing the electricity buy and sell price. Therefore, the optimization problem in the LSE-level can be defined as follows:
% \begin{align}
%     &\max_{C^b, C^s}
%     \begin{aligned}
%         \mathcal{R}^{LSE} \label{eq:LSEobj}
%     \end{aligned} \\
%     &\text{Subject to:} \notag\\
%     &\eqref{eq:GenLimits} \textbf{\&} \eqref{eq:Wind constraint} \textbf{\&} \eqref{eq:GenRamp} \textbf{\&} \eqref{eq:Powerbalance} \notag
% \end{align}
\begin{equation}
 \textbf{LSE level:} 
 \begin{cases}
 \mathop{\mathrm{maximize}}\limits_{C^b_t, C^s_t} & \mathcal{R}^{LSE},   \\ 
\text{subject to:} & \eqref{eq:Powerbalance} \textbf{\&} 
 \eqref{eq:GenLimits} \textbf{\&} \eqref{eq:Wind constraint} \textbf{\&} \eqref{eq:GenRamp}.  
\label{eq:LSEobj}
\end{cases}
\end{equation}

Here, the objective function includes power procurement from the wholesale market as well as the amount of energy sold to residential users. The constraints include power generation limits, generation ramp rate limits, and the power balance, as previously described.

% %%%%%%%%  LMP price formulation %%%%%%%%%%%%%%%%%%
% \begin{align}
%     &\forall \rho \in arg
%     \begin{aligned}
%         &\left\{ \textbf{min} \sum_{t\in\ T} (\sum \limits_{i = 1}^{\mathcal{N_G}} P_{i}^{G}(t)C_i(t))+P^{wp}(t)C_{wp}(t) \right \} \label{eq:LMP}
%     \end{aligned}\\
%     &\text{Subject to:}\; \eqref{eq:GenLimits}\; \text{\&} \; \eqref{eq:GenRamp} \notag\\
%     & \sum \limits_{i = 1}^{\mathcal{D}} L_i(t)   = \sum \limit_{i=1}^{\mathcal{N_G}} P_{i}^{G}(t) + P^{wp}(t)\label{eq:Powerbalance}
% \end{align}

% \subsection{Optimization Problem}
% According to the problem formulation so far, we have a two-level optimization problem. On one hand, the LSE participate in wholesale market to procure the energy needs for residential users, while trying to maximize its local profit. On the other hand, in the residential users level, active prosumers aim to reduce their electricity bills. Given these two observations, the optimization problems can be defined as follows.

% The power balance equation is provided in \eqref{eq:Powerbalance}, also state of charge of the battery, battery charging and discharging limits, and PV panel generation limit are provided in \eqref{eq:batttery_limit} to \eqref{eq:PVconstraint}, respectively. According to \eqref{eq:Powerbalance}, $&P_{k}^{pro}$ can take positive or negative values depending on the amount of PV generation and battery charge or discharge action.
% % Equivalent optimization problem with difference of forecasting and actual

\section{Integrated LSTM and DDPG Framework}\label{DRLframework}
\begin{figure*}
   \centering
    \includegraphics[width = \textwidth]{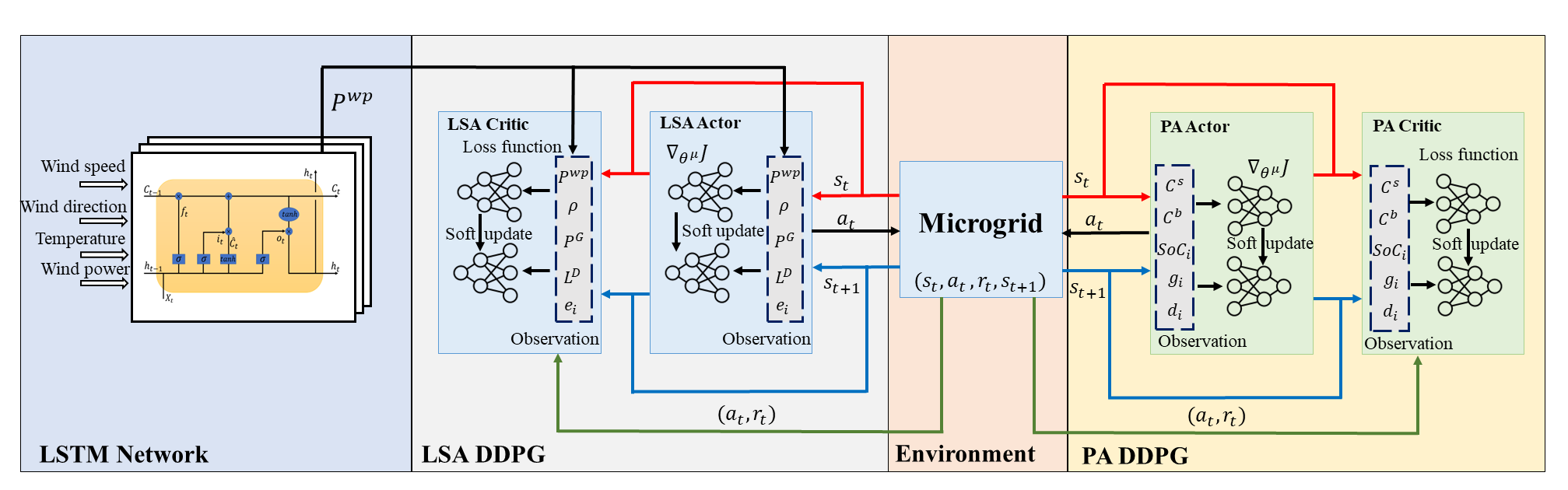}
    \caption{Our proposed framework that integrates LSTM forecasting with DDPG agents. The LSE DDPG agent determines the electricity price based on the observation vector, and the DDPG agents on the prosumer side take action on battery charging/discharging. }
    \label{Flow}
    \vspace{-.15in}
\end{figure*}

To tackle the formulated optimization problems, we develop a framework consisting of a multi-agent DDPG algorithm integrated with LSTM, as shown in Fig.~\ref{Flow}. This framework consists of two types of DDPG agents: (1) Load Serving Agent (LSA) that is located at the LSE level, and (2) Prosumers' Agents (PAs) that are located at the prosumer level. Furthermore, an LSTM forecasting engine is integrated into the LSA to combat wind power uncertainty. In this case, the LSTM forecasts the next 24-hour wind generation based on the collected data, and supports the LSE participation in the day-ahead market.
% In detail, at any given time, first, the LSTM engine forecasts the next 24-hour of the wind generation based on the collected data of the most correlated features to support LSE by participating in day-ahead market. 
% After that, the LSA observes the prediction and uses the data supplied by the determined LMPs and residential users network to determine the retail electricity sell and buy prices.
Then, the LSA observes this prediction alongside the information from the determined LMPs and residential users network to determine the electricity sell and buy prices for the retail market.
In response to the price signals set by the LSA, the PA decides whether to support the LSE or not by taking actions in terms of amount of battery charge and discharge. To better analyze the proposed LSTM-DDPG solution for sequential decision-making, in the remainder of this section, we discuss the background, architecture, training and validating procedure of the proposed framework. 
% Finally, the PA decides whether to support the LSE or not based on the electricity prices by declaring the action on the amount of energy charge/discharge of their batteries.

\vspace{-0.1in}
\subsection{Background}
\textbf{Long-Short Term Memory:}
LSTM is an enhanced version of recurrent neural network (RNN). LSTM is proposed in~\cite{hochreiter1997long} as a possible solution to overcome the major weakness of RNNs, which is handling time-series data with long-range time dependencies. LSTM cell includes a \textit{memory cell} that can maintain information in memory for long periods of time. As a result, this memory cell allows the LSTM to learn longer-term dependencies in the time-series, and makes it an appropriate choice for time-series forecasting. An LSTM cell consists of three gates that control the flow of information within the LSTM cell. These three gates are: (i) an input gate, (ii) an output gate, and (iii) a forget gate.
To memorize the sequential information of data, LSTM back propagates the gradient of output with respect to input from the end to the beginning.  In our model, we stack several LSTM layers to enhance forecasting accuracy. 
% We developed a deep LSTM network for time-series forecasting application. In the developed DLSTM, we stacked several LSTM layers to obtain accurate forecasting.

\textbf{Deep Deterministic Policy Gradient:}  Deep RL (DRL) frameworks are effective solutions for handling sequential decision-making problems. DDPG is a model-free, off-policy, gradient-based RL framework, which combines the DPG method introduced by Silver in 2014~\cite{silver2014deterministic} and DQN. Similar to the standard RL methods, DDPG framework can be described by a five-tuple $\{\mathcal{S},\mathcal{A},r, p,\gamma\}$, in which $\mathcal{S}$ and $\mathcal{A}$ are the set of states and actions; $r$ is the reward function, and $p$ is a transition function between different states. DDPG consists of two networks: Actor and Critic, where the parameterized actor function $\mu(s|\theta^{\mu})$, with parameter $\theta^{\mu}$, holds the policy and deterministically maps the states to a specific action. The term Deterministic refers to the fact that the actor network provides an exact output instead of a probability distribution over actions, $\mu(s) = \arg\max_a{Q(s,a)}$. In addition, the critic network describes as action-value function $Q(s,a|\theta^{Q})$, with parameter $\theta^{Q}$. In order to facilitate training and guarantee convergence, DDPG creates a copy of these two networks: actor target with parameter $\theta^{\mu^\prime}$, and critic target with parameter $\theta^{Q^\prime}$.
The learning process of DDPG consists of two phases. Similar to Q-learning, the critic estimates the $Q$-values using the Bellman equation:
\begin{align}\label{eq:Bellman}
Q(s,a) = r + \gamma {\mathbb E}_{s^\prime \sim s, a^\prime \sim \pi}[Q(s^\prime,a^\prime)],
\end{align}
where $\gamma \in [0,1]$ is the discount rate, and $(s^\prime,a^\prime)$ denotes the next state-action pair. In DDPG, the next-state Q-values are calculated with the target value and target policy network. Thus, the Critic estimates $Q$-values and updates its parameters by minimizing the mean-squared loss function between the updated $Q$-value and the original $Q$-value, as follows:
\begin{align}\label{eq:LOSS}
L(\theta^{Q})={\mathbb E}_{(s,a,r,s^\prime)\sim\mathcal{B}}[(r+\gamma Q^\prime (s^\prime,\pi^\prime(s^\prime)-Q(s,a|\theta))^2],
\end{align}
where $\mathcal{B}$ denotes the replay buffer. For the policy function, the goal of the actor is to maximize the expected discounted returns by interacting with the environment as  $J(\mu) \approx {\mathbb E}[Q(s,a)|s=s_t, a_t=\mu(s_t)]$. To calculate the policy loss, we take the derivative of the objective function $\nabla_{\theta^{\mu}}J(\mu) \approx {\mathbb E}_s[\nabla_{\theta^{\mu}}Q(s,a)]$, respect to the policy parameter. By applying the chain rule, the actor is updated as follows:
\begin{align}\label{eq:DDPG}
\nabla_{\theta^\mu}J\approx \frac{1}{N}\sum_{i\in \mathcal{B}}{\nabla_aQ(s,a)|_{s= s_i,\ a=\mu(s_i)}\nabla_{\theta^\mu}\mu(s|\theta^\mu)}|_{s= s_i}.
\end{align}
In order to minimize the loss function in Eq.~\eqref{eq:LOSS}, DDPG calculates the gradient of $L(\theta^{Q})$ and updates the critic network parameters by gradient descent as follows:
\begin{align}\label{eq:updateCritic}
\theta^{Q} \leftarrow \theta^Q - lr_c\nabla_{\theta^Q}L(\theta^Q), 
\end{align}
where $lr_c$ is a small learning rate. On the other hand, to improve the performance of the policy and maximize the accumulative return. The actor network updates its parameters using gradient ascent with a small learning rate $lr_a$ as follows:
% by $\theta^{\mu} \leftarrow \theta^\mu + lr_a\nabla_{\theta^\mu}J$.
\begin{align}\label{eq:updateActor}
\theta^{\mu} \leftarrow \theta^\mu + lr_a\nabla_{\theta^\mu}J. 
\end{align}

After updating $\theta^\mu$ and $\theta^Q$, DDPG softly updates the target critic and actor networks with a small constant $\tau$.
\begin{align}\label{eq:updateDDPG}
\theta^{Q^\prime} \leftarrow \tau \theta^Q + (1-\tau)\theta^{Q^\prime},\ \ \
\theta^{\mu^\prime} \leftarrow \tau \theta^\mu + (1-\tau)\theta^{\mu^\prime},
\end{align}
where $\tau\ll 1$ greatly improves the learning stability. Given the system model proposed in Fig.~\ref{Model}, our DDPG agents interact with the electricity market environment to maximize their local returns from the environment, as illustrated in Fig.~\ref{Flow}. In the following sections, we first describe the environment and then present the DDPG agents by defining their actions and reward functions.

\subsection{Environment Setup}
The dynamic energy market proposed in Fig.~\ref{Model} is considered as our RL environment. 
% which is characterized by certain states and reward functions. The rewards describe the evolution of the states given the actions chosen by the agents. 
In this case, the PA and LSA agents interact with the environment and their goal is to gather maximum reward possible from the environment through their actions. Each agent is only able to observe a subset of  environment states due to various factors such as physical limitations, and privacy and data security. We denote all the states of the environment by $\mathcal{S}=  \mathcal{S}^{PA} \cup \mathcal{S}^{LSA} $, wherein $\mathcal{S}^{PA}$ and $\mathcal{S}^{LSA}$ denote the set of observable states by the PA and LSA agents, respectively. 

\textbf{PA States}: At each time slot $t$, the PA state for the $i^{th}$ prosumer is defined as $$s^{PA}_{i,t} = \{ d_{i,t}, g_{i,t}, SoC_{i,t}, C^s_{t-N}, ..., C^s_{t}, C^b_{t-N}, ..., C^b_{t}\} \in \mathcal{S}^{{PA}},$$
% where $d_{i,t}$, $g_{i,t}$, and $SoC_{i,t}$ are the $i^{th}$ prosumer power consumption, PV generation, and battery state of charge respectively. 
% Furthermore,
where $(C^s_{t-N}, ..., C^s_{t})$ and $(C^b_{t-N}, ..., C^b_{t})$ denote the electricity sell and buy price in the past $N$ steps. For the PA, the future price is uncertain, thus observing the past steps of the electricity price helps the PA to infer future price trends as suggested in~\cite{8521585}. 

\textbf{LSA States}: At the same time step $t$, the LSA state is defined as 
% \begin{align*}
%     s^{LSA}_t = \{ \Tilde{P}_{t}^{wp},... ,\Tilde{P}_{t+24}^{wp},{\bf{\rho}}_{t-m}^{DA}, \\ ..., {\bf{\rho}}_t^{DA}, {\bf{\rho}}_{t-m}^{RT}, ..., {\bf{\rho}}_t^{RT}, {L_{t}^{DA}}, {L_{t}^{RT}}, {E_{t}}\} \in \mathcal{S}^{LSA}
% \end{align*}
\begin{equation*}
\begin{split}
s^{LSA}_t = & \{ \Tilde{P}_{t}^{wp},... ,\Tilde{P}_{t+24}^{wp},{\bf{\rho}}_{t-M}^{DA},  ..., {\bf{\rho}}_t^{DA},
{\bf{\rho}}_{t-M}^{RT}, ..., {\bf{\rho}}_t^{RT},  \\ 
& {L_{t}^{DA}}, {L_{t}^{RT}}, {E_{t}}\} \in \mathcal{S}^{LSA}
\end{split}
\end{equation*}
where $\Tilde{P}_{t}^{wp},... ,\Tilde{P}_{t+24}^{wp}$ are the predicted wind generation over the next 24 hours, which comes from the LSTM engine. As described, the uncertainty generated from wind power shifts to the uncertainty in LMP prices. Thus,
% to better help the PA to capture the future price trends,
the LSA observes the last $M$ time step day-ahead and real-time LMPs, denoted as ${\bf{\rho}}_{t-M}^{DA}, ..., {\bf{\rho}}_t^{DA}$ and ${\bf{\rho}}_{t-M}^{RT}, ..., {\bf{\rho}}_t^{RT}$. Moreover, ${L_{t}^{DA}}$ and ${L_{t}^{RT}}$ are the network demand in day-ahead and real-time markets, respectively. ${E_{t}}$ denotes the total amount of power purchased from the prosumer network at time $t$, which is calculated as ${E_{t}}=\sum_{i = 1}^{|\mathcal{N^P}|} (1-\Tilde{\lambda}_i)|e_{i,t}|$.

\textbf{State Normalization:} As described in the previous section, the environment states consist of several features, which have different ranges and distributions. To make all the variables contribute equally,  at any given, we standardize the set of the observations before feeding them to the neural networks. This is known as state normalization. In addition, for the layers' normalization, we use Batch Normalization (BN) to improve the Lipschitzness of the loss function and increase the stability and predictability of the gradients, which decreases the gradient vanishing problem and improves the training speed~\cite{santurkar2018does}. 

% BN first time suggested in~\cite{ioffe2015batch} for reducing the Internal Covariant Shift (ICR), which refers to the change in the distribution of the layers' input. But later in~\cite{santurkar2018does}, it realized that BN improves the Lipschitzness of the loss function and increase the stability and predictability of the gradients. Thus, we leverage BN to decrease the gradient vanishing problem, which improved our training speed.
\subsection{DDPG Agents Setup}
The PA and LSA agents interact with the environment by performing an action iteratively to maximize their local long-term returns. At any given time slot $t$, each agent observes its corresponding observable states $s_t$ from the environment and takes an action $a_t$. After that, the environment provides the agent an immediate reward $r_t$, reflecting the benefits of the action, and transitions to the next state $s_{t+1}$. Thus, each agent independently learns its own policy.

\textbf{Prosumer Agent (PA) Setup:} For the $i^{th}$ prosumer, the charge/discharge command to the energy storage is the action determined by the $PA_i$, which is shown by $a^{PA}_{i,t}=\{ b_{i,t}\} \in\mathcal{A}^{PA}_i$, such that $b_{i,t}$ is continuous action in $[b_t^{min},b_t^{max}]$.
% $\mathcal{A}^{PA}_i$ is the finite set of available actions to $PA_i$.
The ultimate goal of the PA is to minimize the local billing cycle in Eq.~\eqref{eq:prosumerobj}, which is defined as follows:
\begin{equation}\label{eq:PA_reward}
r^{PA_i}_{t} = \sum \limits_{t\in T} \{  (1-\Tilde{\lambda}_i) e_{i,t}C^s_t +  \Tilde{\lambda}_i e_{i,t}C^b_{t} \},
\end{equation}
where $r^{PA_i}_{t}$ represents the $i^{th}$ prosumer reward.
% and $\lambda_i \in \{0,1\}$ in which $\lambda_i=1$ when $e_{i,t}< 0$ and $\lambda_i=0$ when $e_{i,t} \ge 0$.

\textbf{Load Serving Agent (LSA) Setup:} To solve the optimization problem in Eq.~\eqref{eq:LSEobj}, at any given time, the LSA determines the electricity sell and buy prices $C_t^s$  and $C_t^b$ to control the energy management over the network. Therefore, the action space for the LSA is shown as $a^{LSA}_t= \{ C_t^s,C_t^b\}  \in \mathcal{A}^{LSA}$, where $C_t^s$ and $C_t^b$ are continuous actions in $[C_t^{min},C_t^{max}]$. 
% where $\mathcal{A}^{LSA}$ is the finite set of available actions for LSA, i.e., all the possible electricity sell and buy prices. 
To maximize the profit of the LSA in Eq.~\eqref{eq:RevLSE}, the LSA reward function is defined as follows:
\begin{equation}\label{eq:LSA_reward}
r^{LSA}_t = L^D_t C^s_t - \sum\limits_{i = 1}^{{|\mathcal{N^P}|}} {\Tilde{\lambda}_i |e_{i,t}|C^b_t} - (P^G_t + P^{wp}_t)\rho_t,
\end{equation}
where $r^{LSA}_t$ is the LSA reward at time slot t. 

\begin{algorithm}[t]
\caption{The Proposed DDPG and LSTM Training Pipeline}
\label{alg:algorithm}
Train the LSTM engine using historical data\\
Generate day-ahead predictions using LSTM \\
\textbf{Algorithm:}
\begin{algorithmic}[1]
    \FOR{each DDPG agent}
    \STATE \textbf{Initialize} critic, actor networks $Q(s,a|\theta^{Q})$, $\mu(s|\theta^{\mu})$ with parameters $\theta^Q$, and $\theta^\mu$\\
    \STATE \textbf{Initialize} target critic, actor networks with parameters $\theta^{Q^\prime} \leftarrow \theta^Q$, and $\theta^{\mu^\prime} \leftarrow \theta^\mu$\\
    \STATE \textbf{Initialize} replay buffer $\mathcal{B}^{PA}$ and $\mathcal{B}^{LSA}$
    \ENDFOR
    \FOR{each Episode}
        \STATE \textbf{Initialize} the environment, set $s^{PA}_{.,0} \rightarrow 0 $, $s^{LSA}_{0} \rightarrow 0 $\\
    \FOR {each iteration $t \in T$}
        \STATE LSA determines the $C_t^b$ and $C_t^s$ in Eq.~\eqref{eq:LSEobj}\\
        \STATE LSA broadcast the electricity sell/buy prices to minimize Eq.~\eqref{eq:RevLSE}\\
    \FOR {each prosumer $k \in \mathcal{N}^{\mathcal{P}}$}
        % \state Observe prosumer $k$ observable states $s^{PA}_{k,t}$\\
        \STATE $PA$ determines $b_{k,t}$ to minimize Eq.~\eqref{eq:UserLevel}, then create $s^{PA}_{k,t+1}$\\
        \STATE PA receives the corresponding reward $r_t^{PA_k}$
    \ENDFOR
    \STATE PA broadcast the new prosumers' network profiles $L_{t+1}^{DA}$, $L_{t+1}^{RT}$, $E_{t+1}$ back to LSA\\
    \STATE LSA creates $s^{LSA}_{t+1}$ and receives corresponding reward $r_t^{LSA}$ \\
    \IF{$t \leq T_{cap}$}{
        \STATE Store $(s^{PA}_{k,t},b_{k,t},r_t^{PA_k},s^{PA}_{k,t+1})$ in $\mathcal{B}^{PA}$\\
        \STATE Store $(s^{LSA}_{t},{C_t^b,C_t^s},r_t^{LSA},s^{LSA}_{t+1})$ in $\mathcal{B}^{LSA}$}
    \ELSE{
        \STATE Randomly choose mini-batch tuples from $\mathcal{B}^{PA}$ and $\mathcal{B}^{LSA}$\\
        \STATE Update LSA and PA with Eq.~\eqref{eq:Bellman}, Eq.~\eqref{eq:LOSS}, Eq.~\eqref{eq:DDPG}, Eq.~\eqref{eq:updateCritic}, and Eq.~\eqref{eq:updateActor}\\
        \STATE Update target networks with Eq.~\eqref{eq:updateDDPG}}
    \ENDIF
    
    \ENDFOR
    \ENDFOR
\end{algorithmic}
\end{algorithm}

\subsection{Integrated LSTM-DDPG Pipeline}
The pseudocode of the training pipeline is given in Algorithm~\ref{alg:algorithm}. 
% As illustrated in Figure \ref{Flow}, the LSTM engine is integrated with the LSA DDPG network to combat the uncertainty generated by wind power. 
As illustrated, 
% the training of LSTM and DDPG networks are independent of each other. In other words, 
in the first phase, we train the LSTM engine using the historical real data and tune it for future utilization. In the second phase, we utilize the trained forecasting engine in training the LSA and PA DDPG agents. For LSTM training phase, as illustrated in Fig.~\ref{Flow}, we use wind speed, wind direction, temperature, and wind power as the inputs of the forecasting engine. These are the most correlated features in forecasting the future wind generation. In order to help LSTM training, all features should be on a similar scale. This helps to stabilize the gradient descent steps. Thus, we normalize the input data to the LSTM. Also, we use Root Mean Squared Error (RMSE) as the performance metric of the forecasting to evaluate the accuracy of our prediction, which is defined as follows:
\begin{equation}\label{eq:RMSE}
    RMSE = \sqrt{\frac{1}{T}\sum_{i=1}^{T}{({P}_{i}^{wp} - \Tilde{P}_{i}^{wp})}^2}.
\end{equation}

Forecasting the future wind helps the LSA DDPG agent in participating in the day-ahead and real-time markets. As discussed in Fig.~\ref{supply}, the mismatch in forecasting leads to mismatch in LMP prices, which affects the LSA's return in Eq.~\eqref{eq:LSA_reward}. Therefore, in the second phase, the future predicted wind is fed to the actor and critic networks of LSA DDPG as observation. At each episode of the training of each DDPG agent, first, we sample a minibatch of the replay buffer $\mathcal{B}$ to evaluate the local and target Q-value. Then, we obtain the gradient of loss function defined in Eq.~\eqref{eq:LOSS} and gradient of policy network defined in Eq.~\eqref{eq:DDPG}. Next, we update the gradients by Eq.~\eqref{eq:updateCritic} and Eq.~\eqref{eq:updateActor}. Finally, we update the target networks by Eq.~\eqref{eq:updateDDPG}.

\begin{figure}[!t]
    \centering
    \includegraphics[scale=0.25, trim = 0.2cm 0.2cm 0.1cm 0.2cm, clip]{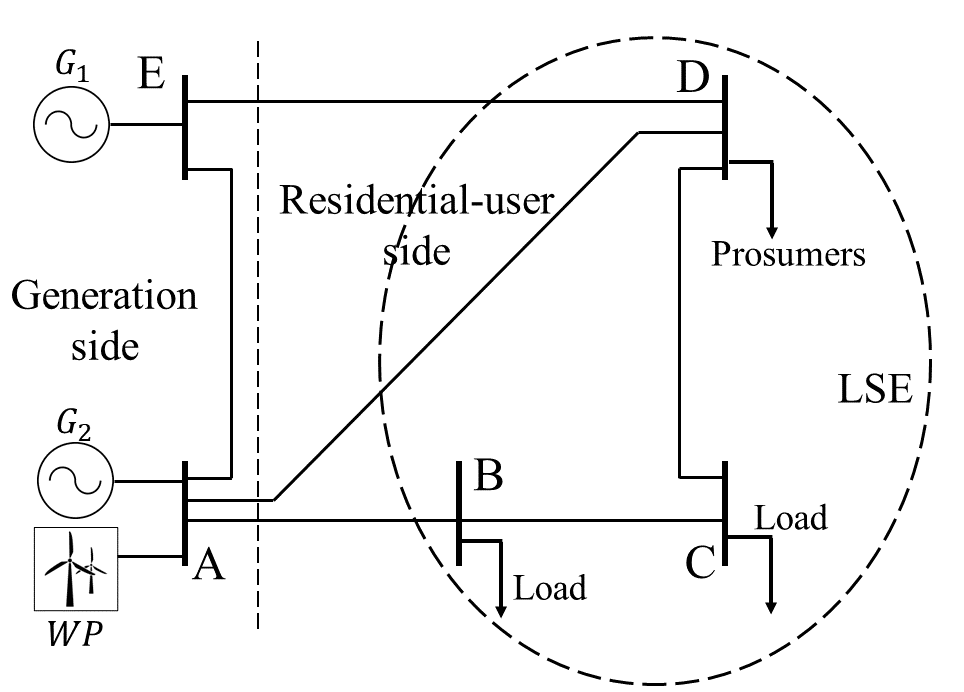}
    \caption{Single-line diagram of the modified IEEE 5-bus system.}
    \label{IEEE5}
    \vspace{-0.1in}
\end{figure}
\section{Numerical Results and Discussion} \label{Numerical}
In this section, we evaluate the performance of the proposed energy management framework in the presence of renewable generation uncertainties. We consider two main sources of uncertainties in renewable generations, one from the wind farm located in the wholesale market, and the other from PV rooftop panels located in the residential network. In addition, to better deal with the uncertainty of the future electricity price, our prosumer agent observes the electricity price history to capture the information in price's trend in real-time. In this section, we first present the experimental settings. Next, we show the PA and LSA behavior, and investigate the performance of the proposed energy management framework in the presence of uncertainties.

\subsection{Simulation Setting}
\textbf{Power system setup:} To evaluate the performance of the proposed energy management framework, we use a modified IEEE 5-bus system, as depicted in Fig.~\ref{IEEE5}. 
% This system is chosen to show how the proposed energy management framework can apply to the standard system models.
On the generation side of this system model, there are two small size gas turbines on buses A and E. Without loss of generality, we forgo the limits for the transmission systems. Generators $G_1$ and $G_2$ are called base and reserve generation units, respectively.
% where the power generated by reserve is more costly. 
In this paper, we consider the quadratic approximation model for the incremental costs for these two generators based on the following cost functions:
\begin{equation}\label{G1:cost}
F(G_1) = \alpha_1 + \alpha_2  P_{1,t}^G + \alpha_3{(P_{1,t}^G)}^2,
\end{equation}
\begin{equation}\label{G2:cost}
F(G_2) = \beta_1 + \beta_2  P_{2,t}^G + \beta_3{(P_{2,t}^G)}^2,
\end{equation}
where the coefficients are derived  from~\cite{ongsakul2013artificial}, and set as $[\alpha_1, \alpha_2, \alpha_3]=[100, 10, 0.2]$ and $[\beta_1, \beta_2, \beta_3]=[200, 15, 0.35]$. The maximum capacity of $G_1$ and $G_2$ are 15MW and 100MW, respectively. In addition, a Wind Power Plant (WPP) is connected to bus A with the price 5\$/MWh. The WPP generation profiles are extracted from historical data provided by~\cite{Dataset}. On the residential user side, the LSE performs energy distribution among buses B, C, and D. The prosumers' network is located on bus D, while the loads on buses B and C are the consumers' network. The daily two double-peak load and PV generation curves mimic the real-world trends reported by California ISO to exemplify real-world operation~\cite{9654550}. 

\textbf{LSTM and DDPG setup:} As described in Section~\ref{DRLframework}, we implement DDPG agents for the LSE and PAs. To do that, the hyper-parameters and simulation setups used for DDPG agents are listed in Table~\ref{tab:SimulationParam}. The simulation setup is implemented in Python with PyTorch 1.12.1. Simulation results are obtained via episodic updating across 4000 episodes, each of which represents a 24-hour cycle with the sampling time of  15 minutes. 
\begin{table}[t]
\resizebox{\columnwidth}{!}{
\centering
\footnotesize
  {
    \begin{tabular}{l|l|l}
      \hline
      \textbf{Hyperparameters} & \textbf{Value for $PA$} & \textbf{Value for $LSA$}\\
      \hline
      Batch size & 64 & 100 \\
      Discount factor & ${\gamma}$=0.95 & $\gamma$=0.95 \\
      Actor/Critic Optimizer & SGD/AdamW & SGD/AdamW\\
      Actor Learning Rate/Momentum & 5e-4/0.8 & 3e-5/0.9\\
      Critic Learning Rate & 5e-3 & 3e-4\\
      Target Smoothing & $\tau$=0.005 & $\tau$=0.005 \\
      Layers/Nodes & 4/[1000,1000,500,1] & 4/[1000,1000,500,1] \\
      Actor Activation Functions & [leaky-relu,leaky-relu,leaky-relu,tanh] & [RRelU,RRelU,RRelU,sigmoid] \\
      Critic Activation Functions & [relu,relu,relu,linear] & [relu,relu,relu,linear] \\
      Reply Buffer Size & 1000000 & 1000000 \\
      Training Noise & $\mathcal{N}(0,0.7)$ with decreasing std & $\mathcal{N}(0,0.07)$ with decreasing std \\
      \hline
      \hline
      \hline
      \textbf{Simulation Parameter} & \textbf{Description} & \textbf{Value}\\
      \hline
      $P_1^{G,\min},P_2^{G,\min}$/$P_1^{G,\max},P_2^{G,\max}$ & Max/Min Gas Turbines & 0 ,0 / 15, 100 MW \\
      $P^{wp,\text{max}}$ & WPP Capacity & 50 MW \\
      $SoC^{{\min}}_{i}$ / $SoC^{{\max}}_{i}$  & Max/Min Batteries State of Charge & 10\% / 90\% \\
      $So{C_{{i}}}(0)$ & Initial state of charge & 1 kWh \\
      $Q_i$ & Battery Capacity & 10 kWh \\
      $[{b^{dis,min}_{i}}$,${b^{charge,max}_{i}}]$ & [Min,Max] Allowable Battery & -2/2 KW \\
      & Discharge/Charge Power Range &  \\
      $g^{max}_{i}$ & Max. Allowable Power generation & 7 kWh \\
      & of PV Rooftop Panels &  \\
      $C^s$ & Electricity Price Range & [0.05,0.2] \$/kWh \\
      \hline
    \end{tabular}
    }}
 \caption{HYPERPARAMETERS and SIMULATION PARAMETERS.}
 \label{tab:SimulationParam}
  \vspace{-0.1in}
\end{table}

\begin{figure}[!t]
   \centering
    \includegraphics[scale=0.21, trim = 0.2cm 0.2cm 0.7cm 0.2cm, clip]{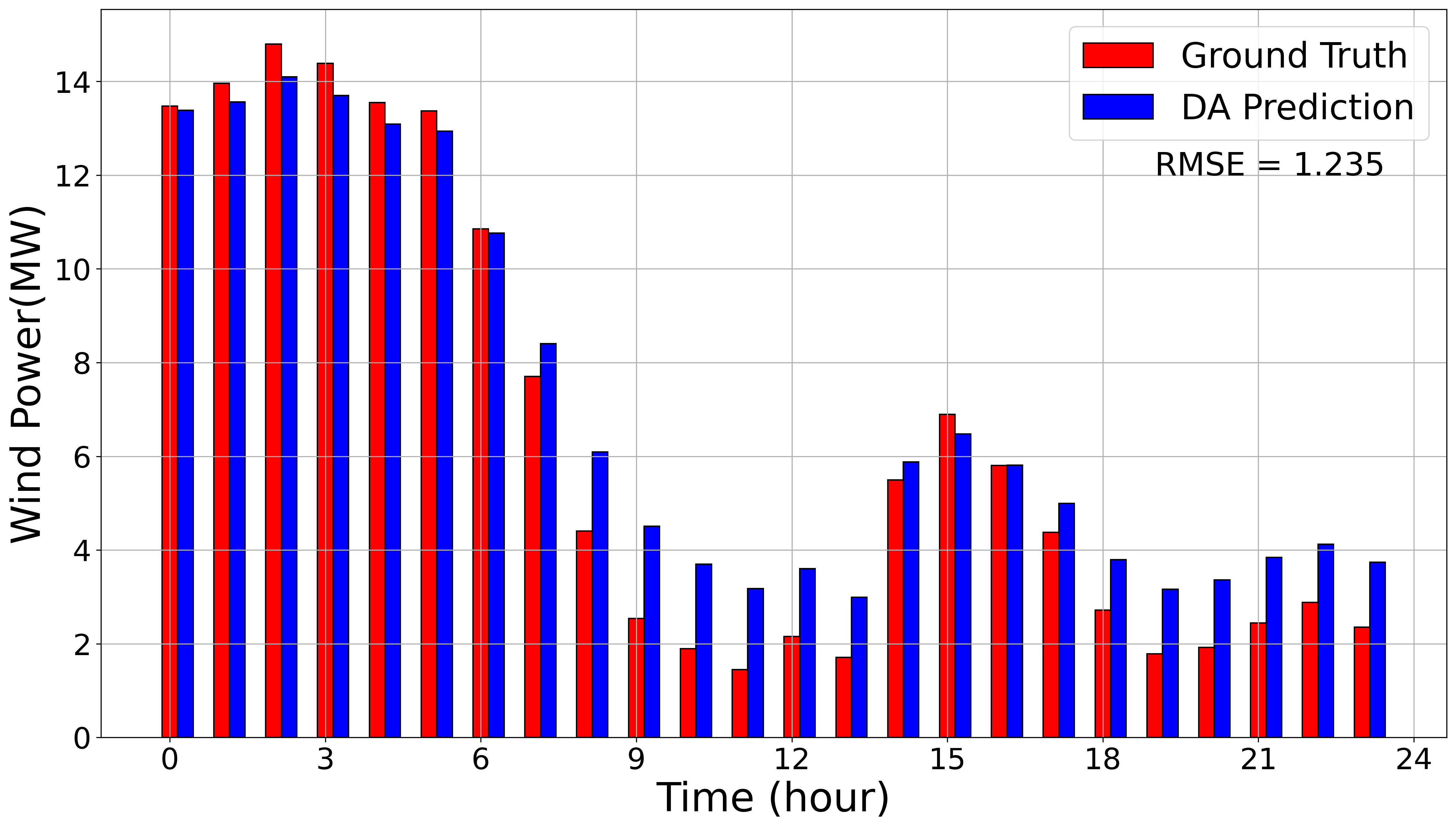}
    \caption{LSTM predication for day-ahead wind power generation.}
    \label{Preds}
    \vspace{-.1in}
\end{figure}

% \subsection{Input Data and Preprocessing}
The wind power data scaled down to fit the test system model. The dataset contains various weather, turbine, and rotor features. This dataset has been recorded for an observation of every 10 minutes from January 2018 till March 2020. 
% Initially, the dataset is preprocessed to visualize the relationship among dataset features.
% A data visualization based on correlation has been demonstrated in the form of a heat map in Fig. \ref{Heat}. 
This initial dataset contains missing values, which could happen due to various reasons, such as faulty recording device, etc. To resolve this issue and improve the performance of our proposed forecasting framework, we use K-nearest neighbor method to estimate the missing values. 
% Since machine learning models are impacted by the quality of data, the dataset has been cleaned and prepared before processing initially, and the missing values in the dataset are imputed using K-nearest neighbor method. 
% After imputing missing values, the data has been visualized to extract the most correlated features. Then, data transformation is implemented, which is an essential pre-processing step that transforms data to the required scale or range. In this regard, the data has been normalized to be between 0 and 1.
% \hl{Add information about the other input parameters}
% Standardization and normalization are various data transformation techniques. Standardization is a data preparation step where numerical values are converted to a common scale without degrading the results. The rescaled data will have a mean of 0 and a standard deviation of 1.
% \subsection{Wind Power Forecasting}
% According to the correlation visualization in data preprocessing, wind speed is the most correlated feature to wind power generation. In this work, we also include wind speed, wind direction, and temperature as the input of the forecasting engine.
In the second step, the dataset is modified to be in a sliding window format with observation of the past 24 hours to develop a short term forecasting of wind power. Next, the data is divided with 80\% for training and the rest for testing the LSTM model. 
% To make the dataset well-structured, the technique, which gives the structured or normalized values within the range of 0 and 1.
In our implementation, a stacked LSTM model consists of 100 neurons with $tanh$ activation function, and the optimizer is set to be Adam with the learning rate of 0.001. The batch size is 64, and the number of iteration epochs is 100.
The forecasting results for day-ahead wind power generation is demonstrated in Fig.~\ref{Preds} for one typical day. From the results, we observe that LSTM is able to accurately predict wind generation over a window of 24 hours. 
% in which the x-axis shows the time in hour and the y-axis represents the power in kW. 

% To illustrate the efficiency of LSTM model in this case, three different methods also has been implemented and the table. \ref{table:Error} shows the comparison for these models.  Fig.\ref{Forecast} shows the forecasting result for next 24 hours.

\subsection{Load Serving Entity Behavior}
The ultimate goal of the LSA is to learn a policy to optimally distribute the electricity and manage the resources among the network,  while increasing its local profit in Eq.~\eqref{eq:LSEobj}. At each time slot, the LSA participates in the day-ahead and real-time markets. In this paper, we assume that the LSA has access to the past wind generation data, but there is no knowledge about the real-time data. Thus, to participate in the day-ahead and real-time markets, the LSA needs an estimation for wind capacity in the near future (i.e., next 24 hours), and thus it uses the LSTM forecasting engine. Further, we set $C_t^b = C_t^s$, which means that the buy and sell prices are equal to model the existing net-metering scenarios. Therefore,  the electricity sell price $C_t^s$ is the only LSA action.

\begin{figure}[!t]
  \centering
    \includegraphics[scale=0.055, trim = 2.8cm 0.2cm 1.2cm 1.5cm, clip]{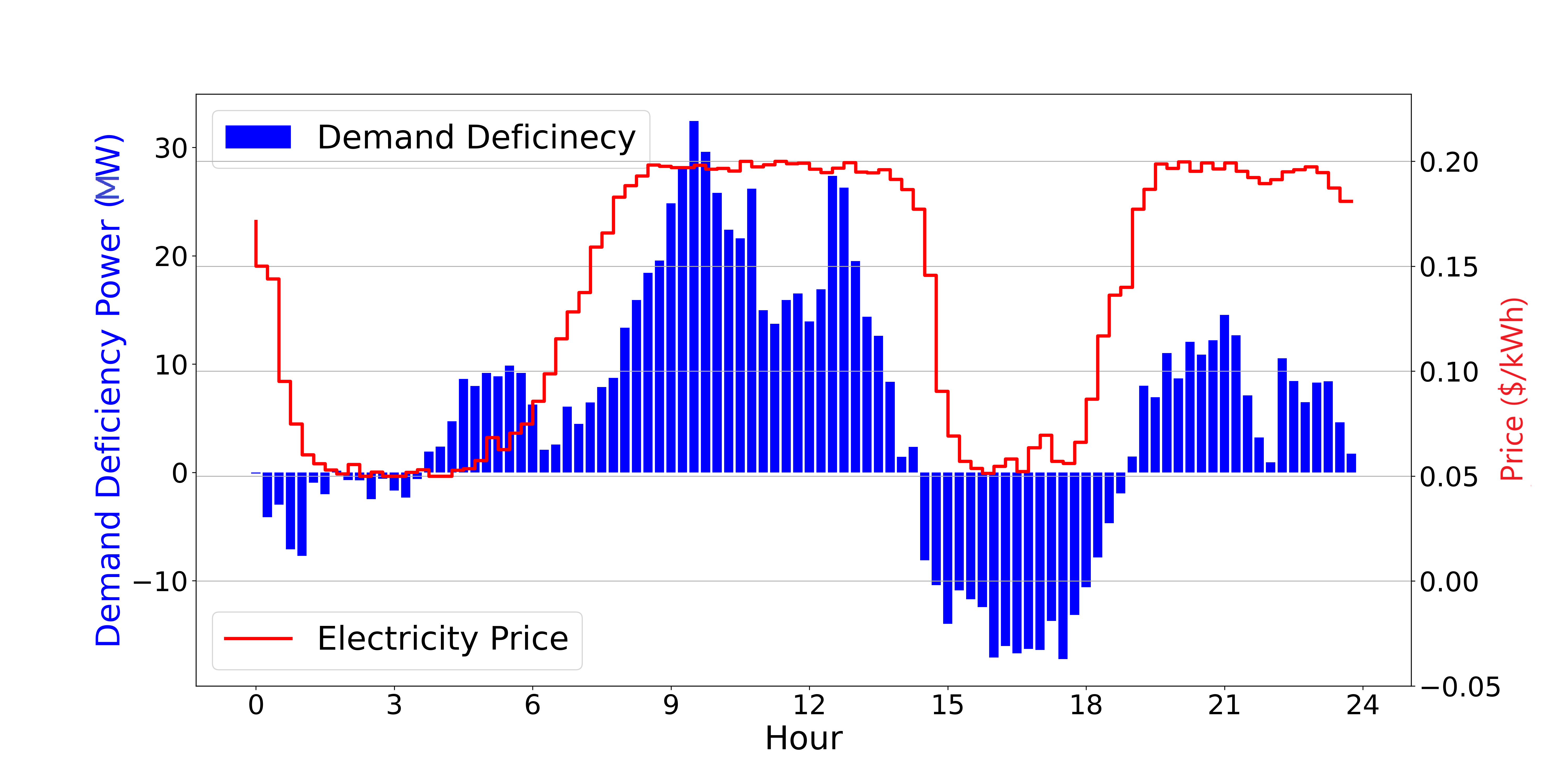}
    \caption{Day-ahead and real-time demand deficiency, and the dynamic pricing scheme generated by the LSA DDPG agent.}
    \label{DemandDeficiency}
    \vspace{-.1in}
\end{figure}

Fig.~\ref{DemandDeficiency} represents a sample day of the behavior of the LSA in real-time. As mentioned, the electricity price is the control variable and action for the LSA. The bar plot in this figure shows the demand deficiency, which can be positive or negative. The positive/negative value for demand deficiency indicates higher/lower real-time demand than what was committed in the day-ahead market. 
% The figure shows that the LSA dynamically changes the electricity price during the day based on the demand deficiency. 
The results demonstrate that the LSA DDPG agent increases the electricity price, when the real-time demand exceeds the amount of energy that the LSA committed to procure in the day-ahead market. As a result, the LSA attempts to compensate for the deficit by procuring from distributed PV sources.
% the user committed in the day-ahead market.

\subsection{Prosumers' behavior}
On the residential-user side, the PA controls the battery charge/discharge command based on the real-time price signal, real-time PV generation, and real-time local consumption. The ultimate goal of the PA is to learn a policy to minimize the total electricity bill in Eq.~\eqref{eq:UserLevel}, in the presence of PV generation and price uncertainties. For the PA, to deal with the uncertainty in future price, we observe the past $N$-step of the electricity price. During extensive simulation, our results show that observing the past 20 steps of electricity price is sufficient to improve the PA decision-making.
To deal with the PV generation uncertainty, we leverage the weather-aware framework proposed in our previous work~\cite{9654550}, by labeling the day-ahead as a $\{Cloudy,Sunny\}$ day and observing the day-ahead label in advance. With this knowledge, our PA would be able to adjust its decision-making behavior to improve the energy distribution over the network with supporting the grid during peak demand hours. 

\begin{figure}[!t]
  \centering
    \includegraphics[scale=0.16, trim = 0.2cm 0.2cm 0.5cm 0.2cm, clip]{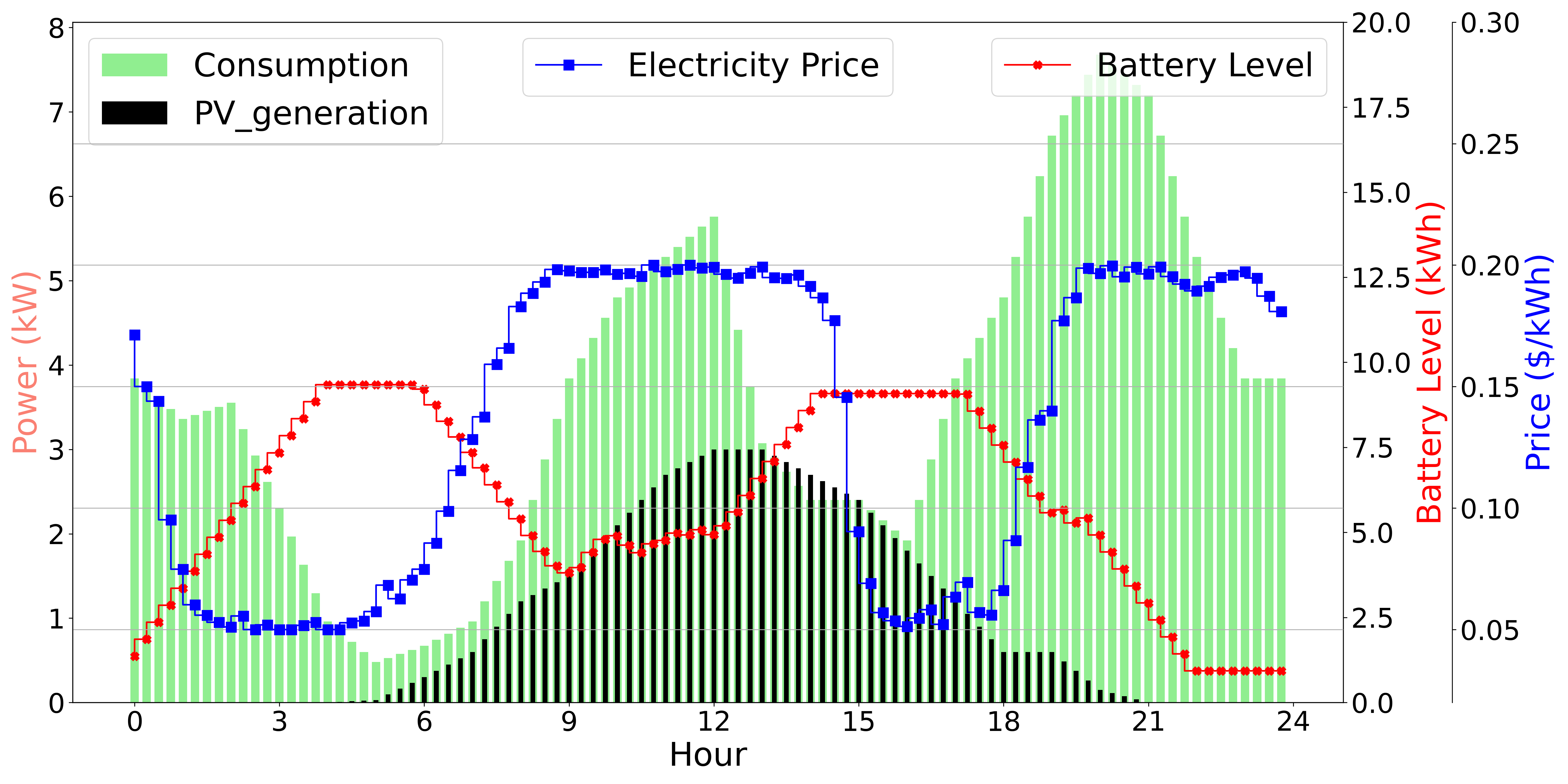}
    \caption{A prosumer profiles of consumption, PV generation, and battery level during a 24-hour window.}
    \label{HouseholdProfile}
    \vspace{-.1in}
\end{figure}

Fig.~\ref{HouseholdProfile} illustrates the behavior of one of the prosumers in its last day of the simulation. This day labeled as a ``cloudy day'' since a small amount of the excess PV generation is discernible. In this day, the agent tries to charge the battery during the off-peak hours when the price is relatively low, especially, at the beginning of the day when there is no excess PV power to support the grid during the peak demand hours. The behavior of purchasing during off-peak demand and selling back to the grid during the peak-demand hours is also called battery \textbf{arbitrage}, which not only supports the grid to decrease its peak load, but also increases the profit of the prosumers.

\begin{figure}[!t]
  \centering
    \includegraphics[scale=0.16, trim = 1.2cm 0.2cm 0.25cm 0.2cm, clip]{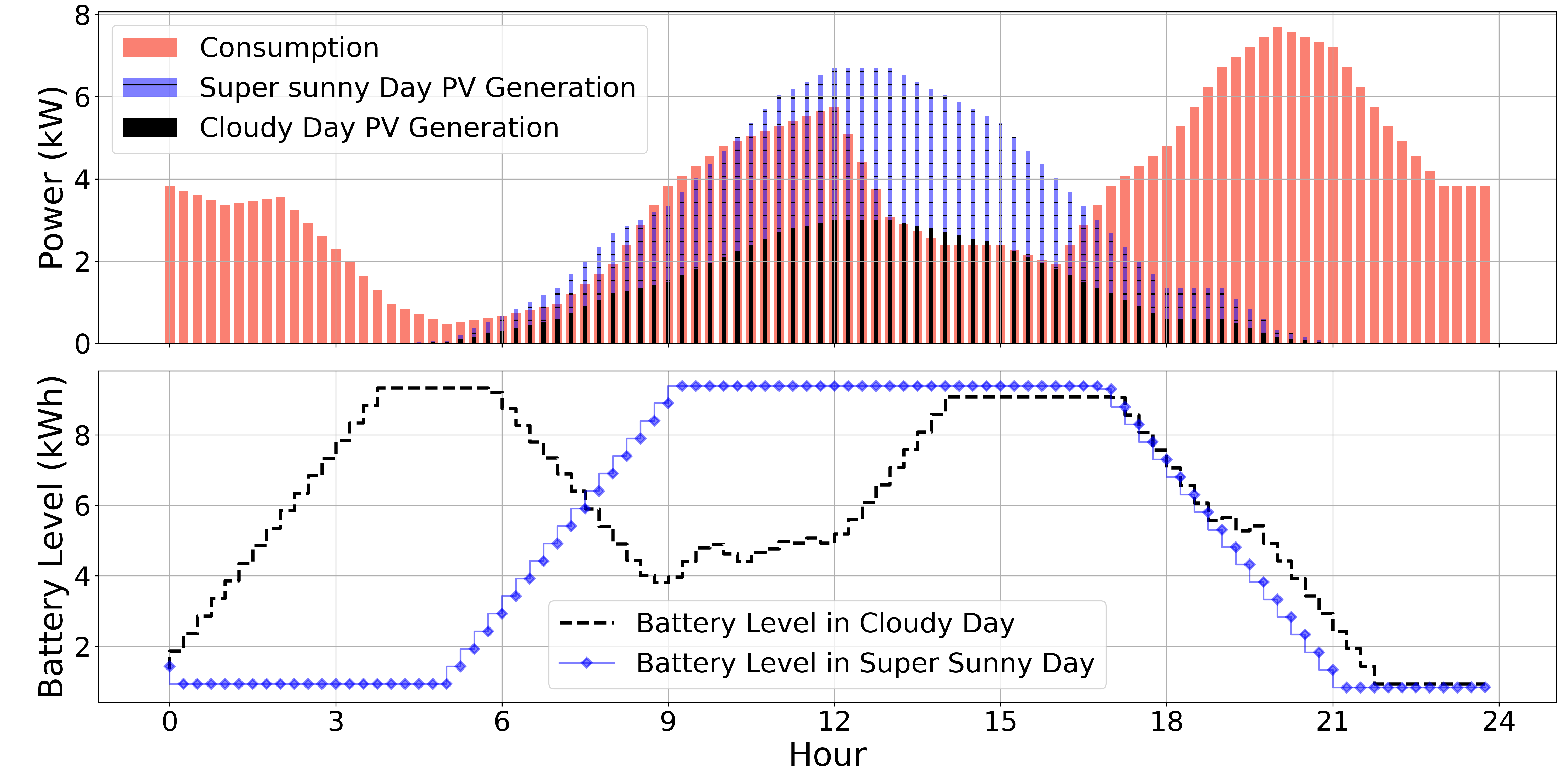}
    \caption{Impact of weather uncertainty on the battery arbitrage.}
    \label{PVuncertainty}
    \vspace{-.1in}
\end{figure}

Fig.~\ref{PVuncertainty} represents how considering the weather uncertainty can affect the behavior of the PA and battery arbitrage. In the case that the day is labeled as a ``sunny day'', the PA prefers to wait for PV excess power to charge the battery with excess power and then discharge it during the peak demand hours. Charging with excess power during sunny days ensures higher benefits for the prosumers, compared with purchasing at the beginning of the day.


\subsection{Impacts of Uncertainties}
\begin{figure}[!t]
  \centering
    \includegraphics[scale=0.047, trim = 15cm 0.2cm 4.4cm 0.2cm, clip]{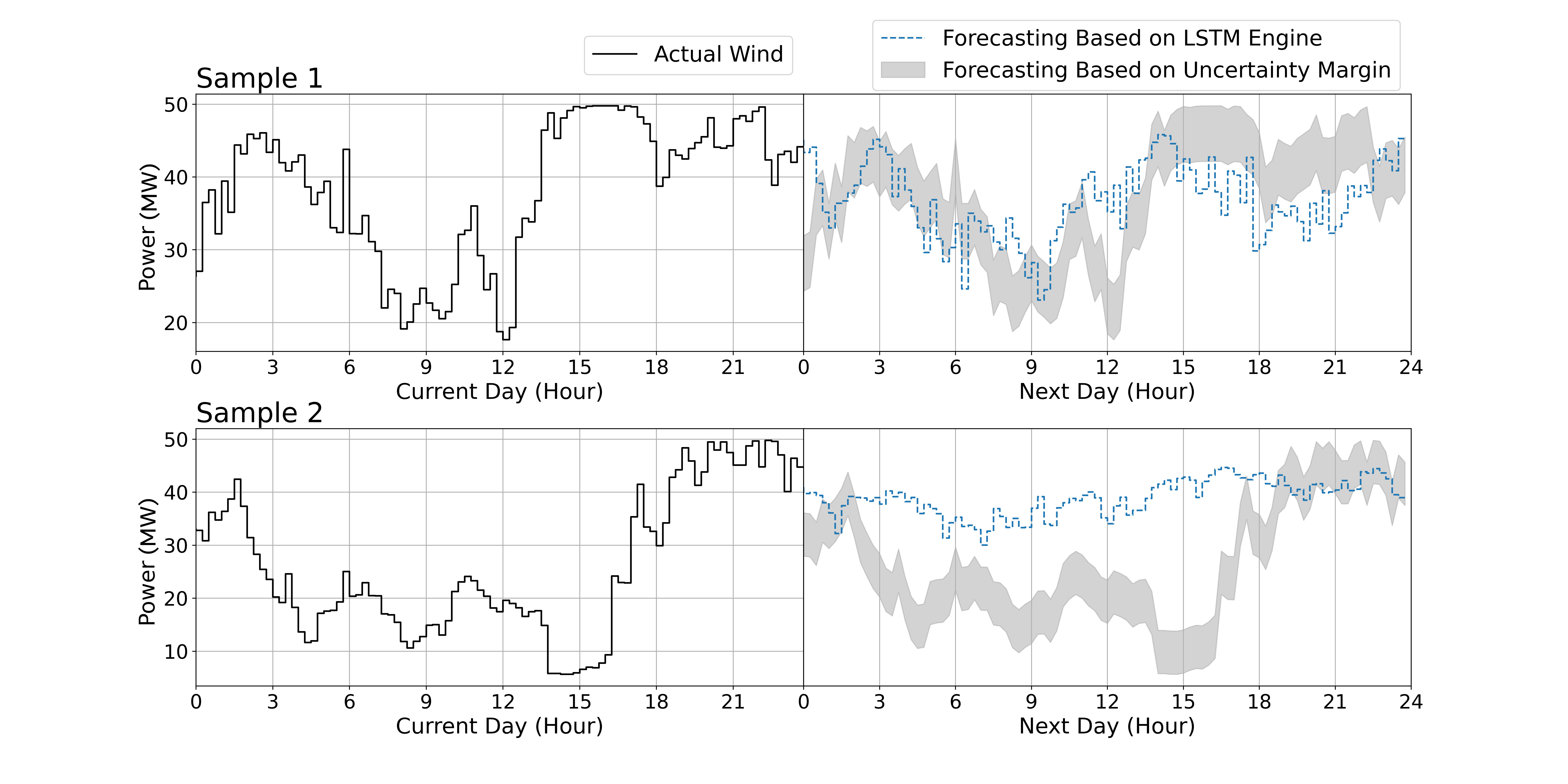}
    \caption{Wind generation for two consecutive days, which are the current day and the day-ahead generation. For the day-ahead prediction, a range of uncertainty is considered as well as forecasting by the LSTM module. }
    \label{ForecastingWaveforms}
    \vspace{-.1in}
\end{figure}

This section investigates the operation of the proposed LSTM-DDPG framework in the presence of renewable generation uncertainty. The results further indicate the importance of considering wind power uncertainty for  decision-making in the real-time and day-ahead markets to optimize the energy distribution in the retail market. To highlight this, we consider two cases: 

\begin{itemize}
    \item \textbf{Case 1} assumes that there is no forecasting engine in the LSE location. In this case, one baseline approach for the LSA is to participate in the DA market by considering an \emph{uncertainty range} for wind generation based on the last 24-hour of real-data, likewise in~\cite{8742669}. In this paper, we consider an $\pm10\%$ uncertainty range, meaning that the forecasted wind generation over the next 24 hours would be within an $\pm10\%$ of the generation over the past 24 hours. 
    \item \textbf{Case 2} considers an LSTM engine in the LSE, which is fully integrated with the DDPG agent. This engine provides the forecasting wind generation in the next 24 hours. Fig.~\ref{ForecastingWaveforms} shows two sample days of the dataset from two different months. 
\end{itemize}

\noindent 
\textbf{(1) Impact of uncertainty on LMP and prosumers.} 
The pattern of wind generation is highly correlated with wind speed, wind direction, and temperature. The differences between intra-day wind generations may be small or large, depending on the wind farm's geolocation.
% It will change during the days. Especially, in windiest areas, it changes more frequently. 
Specifically, if the pattern for the next 24 hours is similar to the previous 24 hours, then obviously there will be a slight difference between the range of uncertainty and forecasted wind, as shown in Sample 1 in Fig.~\ref{ForecastingWaveforms}. Thus, there would be slight mismatch between the real-time and day ahead LMPs. On the other hand, if the wind pattern changes more significantly for the next 24 hours, the difference between uncertainty range and forecasted wind would increase, which in turn, results in a higher mismatch between the day-ahead and real-time LMPs, as it is shown in Sample 2 in Fig.~\ref{ForecastingWaveforms}. In other words, the uncertainty in wind patterns shift to the uncertainty in LMP prices, where it directly affects Eq.~\eqref{eq:CostLSE} and the LSA distribution strategies.
% As pictured in Figure~\ref{Model}, in our system model, there is one LSE that controls the energy distribution over a residential network. 
% As described before, the primary goal of the LSE is to optimally dispatch the distributed energy resources in the residential network to maximize its local profit in the presence of uncertainties. 
% two main sources of uncertainty. One from the wind farm located in powerplant side, which affects the wholesale market. The other source of uncertainty stems from the PV rooftop panels located in residential users, which affects the retail market. 

\begin{figure}[!t]
  \centering
    \includegraphics[scale=0.19, trim = 3cm 2.5cm 3.8cm 4.5cm, clip]{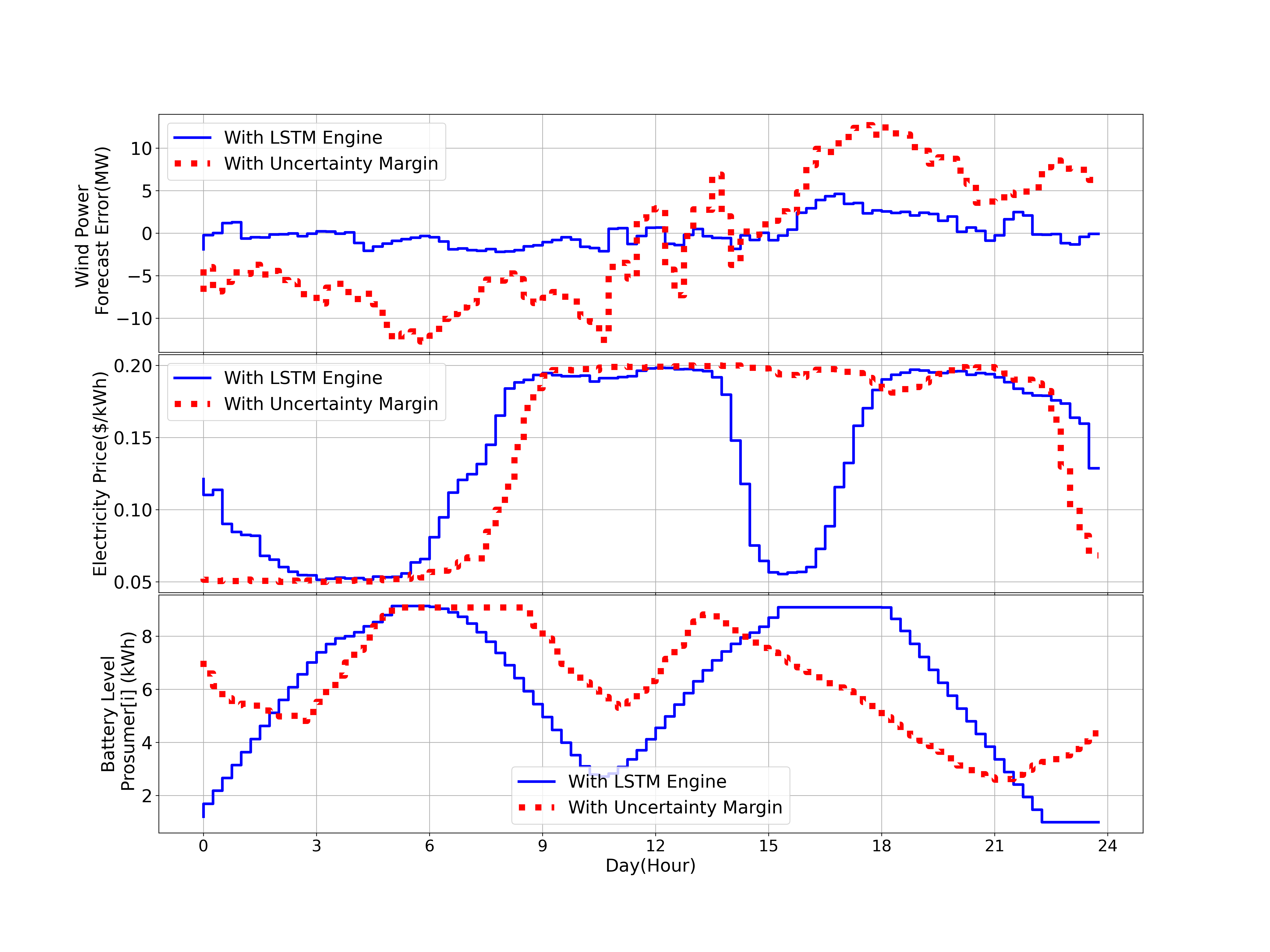}
    \caption{Comparing the prosumer and LSE behavior with the LSTM engine and the method based on uncertainty range.}
    \label{OneDayProfile}
    % \vspace{-.1in}
\end{figure}

Fig.~\ref{OneDayProfile} compares the performance of the two cases in a specific day. As depicted, with uncertainty range, the LSA is not able to dynamically change the price during the afternoon, which directly affects the charging/discharging behaviors of the prosumers. Thus, the prosumers are not able to use the whole span of the battery, which leads to a smaller profit in long-term.

\vspace{.1cm}
\noindent 
\textbf{(2) Impact of uncertainty on Peak-to-Average and LSE profit.} Next, we compare the Peak-to-Average (PAR) performance of the two cases.
PAR ratio has been used extensively in the literature as a parameter to measure the effectiveness of the demand-side management algorithms,
% ~\cite{dewangan2022peak}
and is defined as follows:
\begin{equation}\label{eq:PAR}
\text{PAR} = \frac{T \max_{t \in \mathcal{T}}{L^D_t}}{\sum\limits_{t \in \mathcal{T}}{L^D_t}}.
\end{equation}

\begin{figure}[!t]
  \centering
    \includegraphics[scale=0.19, trim = 2cm 0.2cm 3.2cm 2cm, clip]{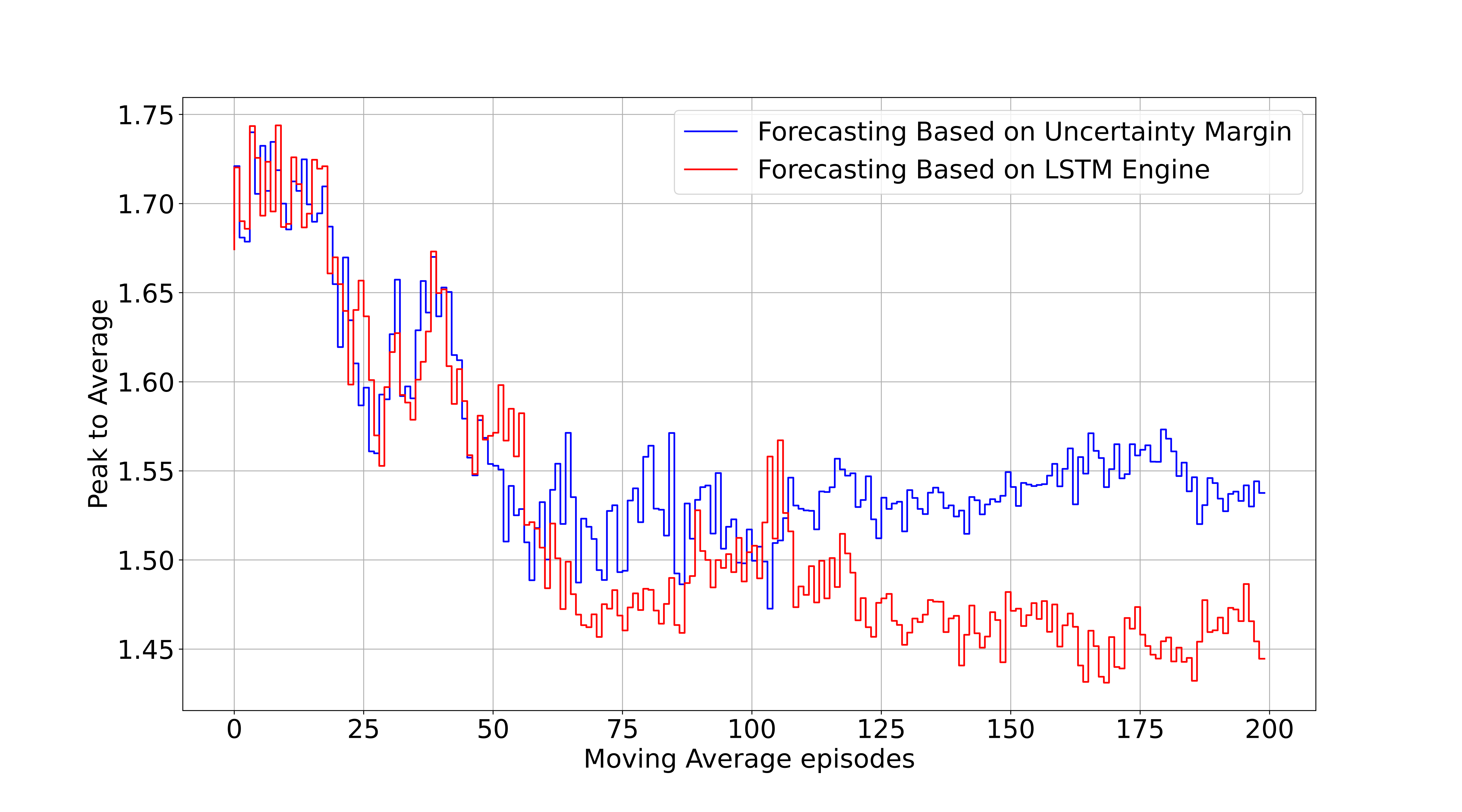}
    \caption{PAR comparison of the proposed method with the scheme based on the uncertainty range.}
    \label{PAR}
    \vspace{-.1in}
\end{figure}

Fig.~\ref{PAR} compares the PAR of the two cases. From the results, we notice that if the LSA integrates the forecasting engine with the DDPG algorithm and observes the predicted wind, this helps the LSA to have a better estimation on what the real-time/day-ahead LMPs would be and this helps the LSA to optimally change the price to incentivize the prosumers to participate in grid support program during peak demand hours. This participation decreases the PAR compared with the case without LSTM forecasting, i.e., using an uncertainty range. 

\begin{figure}[!t]
  \centering
    \includegraphics[scale=0.07, trim = 1.8cm 0.2cm 3cm 1.5cm, clip]{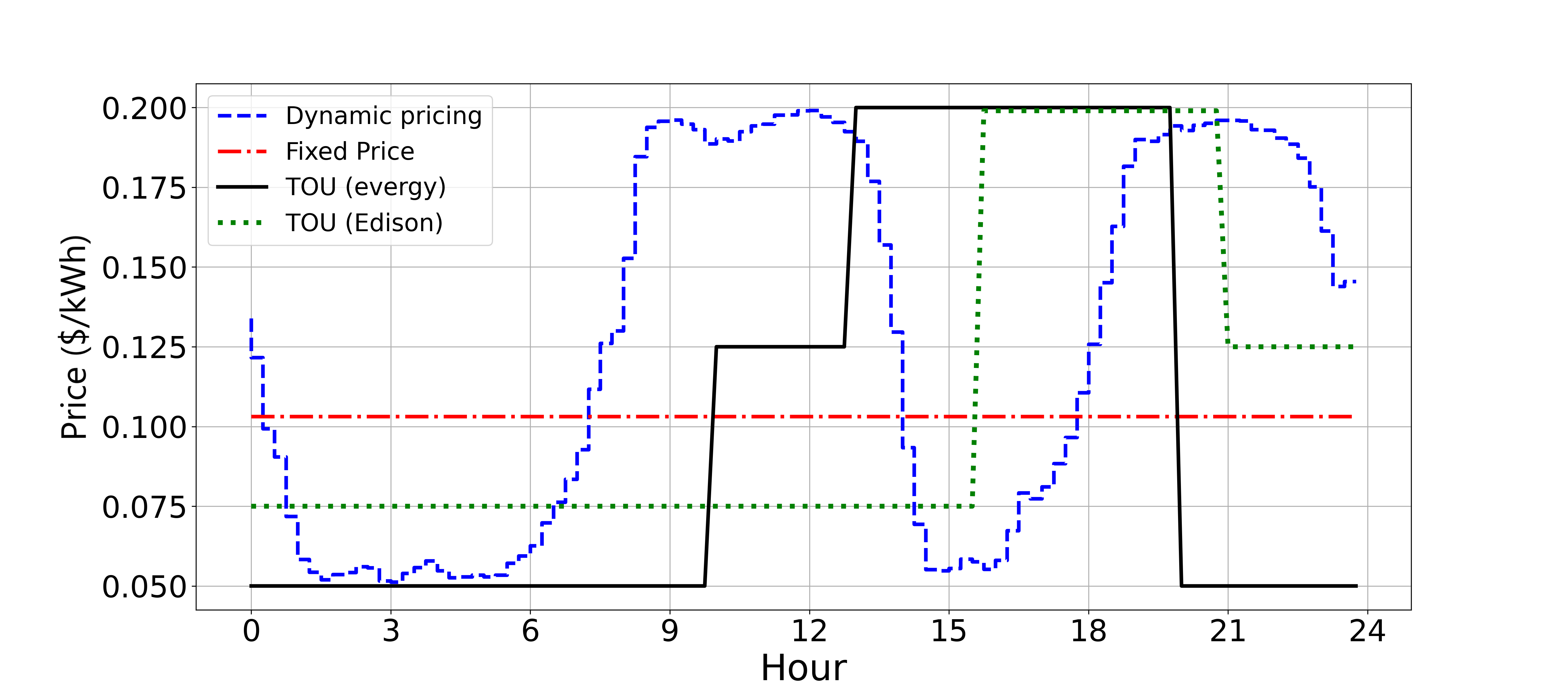}
    \caption{Different electricity price waveforms including, dynamic pricing generated with the proposed DDPG algorithm, a fixed price, and two time-of-use pricing scheme.}
    \label{PriceWaveforms}
    \vspace{-.1in}
\end{figure}

Table~\ref{tab:Scenarios} compares the average PAR and average profit of the last 100 days of the simulation for different pricing scenarios defined in Fig.~\ref{PriceWaveforms}. In this paper, different pricing schemes are compared with our dynamic pricing framework. As shown in Fig.~\ref{PriceWaveforms}, two different time of use (TOU) waveforms have been considered. One of them is from an LSE located in Kansas~\cite{Evergy}, and the other is a waveform used in California~\cite{Edison}. The fixed price waveform is defined as the average of Kansas TOU. Table~\ref{tab:Scenarios} shows that our dynamic pricing based on DDPG framework with an integrated forecasting engine results in higher profits and smaller PAR for the LSE. 
% Even our pricing scheme without forecasting engine outperforms the state-of-the-art TOU and fixed pricing schemes, which are using in different distribution sectors of the US electricity market.

\begin{table}[t]
\resizebox{\columnwidth}{!}{
\centering
\footnotesize
  {
    \begin{tabular}{l|l|l}
      \hline
      \textbf{Senarios} & \textbf{Profit} & \textbf{PAR}\\
      \hline
      Fixed Price & 4.618 & 1.742 \\
      \hline
      TOU (Evergy) & 5.846 & 1.618 \\
      \hline
      TOU (Edison) & 6.211 & 1.605 \\
      \hline
      Dynamic Pricing Based on Uncertainty Margin & 8.536 & 1.542 \\
      \hline
      Dynamic Pricing Based on LSTM Engine & \textbf{10.853} & \textbf{1.462} \\
      \hline

    \end{tabular}
    }}
 \caption{THE LSE PROFIT and PAR PERFORMANCE for DIFFERENT PRICING SCHEMES.}
 \label{tab:Scenarios}
  \vspace{-0.15in}
\end{table}

\section{Conclusion} \label{Conclusion}

In this paper, we studied the impact of renewable energy resources uncertainty in both wholesale and retail markets in smart grid. We formulated the problem as the two-level optimization problem, and developed a framework by combining deep deterministic policy gradient (DDPG) RL and LSTM models. The RL agent for the LSE determines the electricity price, and the prosumer agent determines the battery charge and discharge actions. The LSTM is implemented for time-series forecasting to address the wind power generation uncertainty. Our simulation results demonstrate
that the proposed framework provides higher economic benefits for both LSE and prosumers.
Specifically, properly incentivizing prosumers through dynamic pricing and leveraging the capacity
of distributed battery resources result in: (i) reduced average daily bills for prosumers, (ii) enhanced profits for the LSE by decreasing the reserve generation power demand, and (iii) reduced peak-to-average ratio.

\bibliographystyle{IEEEtranN}
\bibliography{Refs}

\end{document}